\documentclass[%
 reprint,
superscriptaddress,
 amsmath,amssymb,
 aps,
 prb,
]{revtex4-2}

\usepackage{graphicx}
\usepackage[dvipsnames]{xcolor}
\usepackage{dcolumn}
\usepackage{bm}
\usepackage{float}
\usepackage[colorlinks, citecolor=red, linkcolor=blue]{hyperref}

\begin{document}


\title{Multi-qudit interactions in molecular spins}

\author{\'{A}lvaro G\'{o}mez-Le\'{o}n}
\email{a.gomez.leon@csic.es}
\affiliation{Instituto de F\'{i}sica Fundamental IFF-CSIC, Calle Serrano 113b, Madrid 28006, Spain}




\date{\today}

\begin{abstract}
We study photon-mediated interactions between molecular spin qudits in the dispersive regime of operation.
We derive from a microscopic model the effective interaction between molecular spins, including their crystal field anisotropy (i.e., the presence of non-linear spin terms) and their multi-level structure.
Finally, we calculate the long time dynamics for a pair of interacting molecular spins using the method of multiple scales analysis.
This allows to find the set of 2-qudit gates that can be realized for a specific choice of molecular spins and to determine the time required for their implementation.
Our results are relevant for the implementation of logical gates in general systems of qudits with unequally spaced levels or to determine an adequate computational subspace to encode and process the information.
\end{abstract}

\maketitle


\section{Introduction}
Quantum technologies have become one of the corner stones in modern science and engineering. Initially boosted by the prospects created by quantum computers, their range of application continues widening over the years, with influence in the future of communications~\cite{QuantumInternet}, drugs development~\cite{DrugDiscovery} or novel materials with spectacular properties~\cite{NewMaterials}, to name just a few examples.

Although a fully programmable universal quantum computer is still out of reach, we are now entering the era of Noisy Intermediate-Size Quantum devices (NISQs)~\cite{Preskill2018quantumcomputingin}. These devices are designed to perform specific tasks more efficiently than classical computers~\cite{QSupremacy}, and for this reason, they can be fabricated using completely different architectures~\cite{QC-Ions0,QC-Ions1,QC-Ions2,QC-Supercond0,QC-QDots0,BosonSampling}.

One of these architectures is based on magnetic molecules~\cite{QC-Magnets0,FLuis-MolecularSpins&QC,FLuis-ScalableArchitecture,Coronado2020,Vanadyl1,Vanadyl2,Vanadyl3}.
Molecular spins have been studied for some time due to their attractive coherence time and their chemical synthesis control, which allows to design molecules with specific features. 
Lately, molecular spins made of lanthanide ions have attracted a lot of attention~\cite{Lanthanides1,Lanthanides2}, and their integration in hybrid structures offers many possibilities~\cite{Perspective,Strong-coupling0,Strong-coupling1,Strong-coupling2,Strong-coupling-Spin0,Strong-coupling-Spin1,Strong-coupling-Spin2}.
A fundamental one is to consider molecular spins as the building blocks of these hybrid architectures. 
Although this goal requires the coupling of single spins and light~\cite{Spin-Photon0,Vandersypen2018,Benito2019} for the manipulation of isolated qudits and their communication, recent advances seem to indicate that achieving this goal is within experimental reach~\cite{Constriction,Borjans2020,Harvey-Collard2022}.

Crucially, molecular spins are far more complex than the idealized qubits typically considered in quantum computation studies. 
For example, their number of energy levels can be large or their ligand crystal field can introduce non-linear terms that destroy level degeneracies. 
Although these additional features complicate the description of computational tasks, they are not always in detriment of their use as the building blocks of quantum computers~\cite{Qudits&QC,Vandersypen2018}. 
For example, their rich level structure can be used to implement local error-correction codes in each molecule~\cite{ErrorCorrection0,Error-Correction1,Error-Correction2,Error-Correction3}, to define more robust logical qubits~\cite{Qudits0,Qudits1} or to use each molecular spin as a local processor to perform fast logical operations, in addition to the typically slower ones proposed for interacting distant spins. 
Also, it has been shown that the use of qudits offers certain computational advantages~\cite{Qudits-LightMatter,SalesmanProblem}.

\begin{figure}
    \centering
    \includegraphics[width=1\columnwidth]{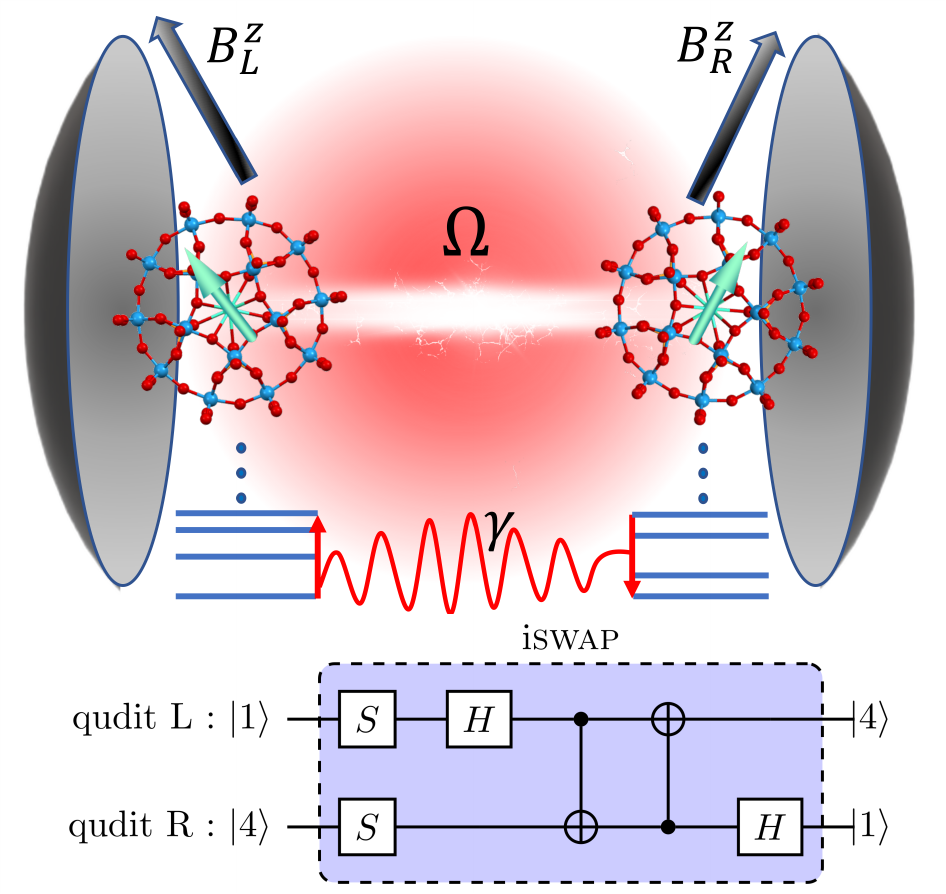}
    \caption{Schematic of two molecular spins of $\text{GdW}_{30}$ controlled by local fields $B_j^z$ and interacting through the cavity photons $\gamma$. Below it is shown the multi-level structure of each qudit and a pair of levels tuned to resonance, which lead to an iSWAP gate between states $|1,4\rangle\leftrightarrow|4,1\rangle$. At the bottom is shown a circuit that would implement the analogous operation.}
    \label{fig:Schematic}
\end{figure}
In this work we derive, from a microscopic model for molecular spins, the effective interaction between qudits in the dispersive regime, mediated by cavity photons.
Our results include a complete description of the molecular spins in terms of Stevens operators and take into account their multi-level structure.
We find that tuning the level splitting in each molecule with local fields allows to control the resonant transitions between spins, via the cavity photons, and therefore to select the 2-qudit operations being implemented (see Fig.~\ref{fig:Schematic} for an schematic comparison of the physics of interacting molecular spins with a particular circuit implementation of logical operations in a quantum computer).
Furthermore, we estimate the time required to implement each gate in order to determine the most accessible set of operations for a particular molecular spin.
We illustrate these results with the analysis of two examples: a pair of N-V centers ($S=1$) and a pair of $\text{GdW}_{30}$ molecules ($S=7/2$)~\cite{GdW-Molecule}.

From a fundamental perspective, this work generalizes previous approaches to effective interactions~\cite{Asenjo-Garcia2019}, by incorporating the role of non-linear terms in the Hamiltonian. This is also incorporated in the treatment of the Schrieffer-Wolff transformation.
Finally, the time evolution is also studied with the method of multiple-scales analysis.
Although this technique is not widely used in the quantum physics community, here it allows to simultaneously deal with all the different resonant transitions and explore all the $2$-qudit gates that can be implemented. This would be far more tedious using perturbative methods only.

\section{Effective Hamiltonian}
We consider a set of qudits interacting with a quantized bosonic field. 
In our case, the qudits correspond to molecular spins, which represent effective descriptions of molecular magnets and can be accurately described by "giant" spins $\vec{S}_i$~\cite{MolecularSpins1,MolecularSpins2}.
As the molecules display crystal anisotropy and a multi-level structure, it is useful to describe them in terms of Stevens operators:
\begin{equation}
    \mathcal{H}_{S}\left(i\right)=\sum_{k=2,4,6}\sum_{q=-k}^{k}B_{k}^{q}\hat{O}_{k}^{q}\left(\vec{S}_{i}\right)+\mu_{B}\vec{B}_{i}\cdot\hat{g}\cdot\vec{S}_{i}\label{eq:Qudit}
\end{equation}
where $\vec{S}_i$ is the spin of the $i$-th molecule, $\mu_B$ is the Bohr magneton, $\vec{B}_i$ the local external magnetic field, $\hat{g}$ the gyromagnetic tensor, $\hat{O}_{k}^{q}$ the extended Stevens operators, which are polynomials of the spin operators, and $B_{k}^{q}$ the corresponding coefficients.
Eq.~\eqref{eq:Qudit} describes our set of qudits, each with $2S_i+1$ unequally spaced energy levels.

The spins are coupled to a quantized bosonic field, which in our case is produced by a superconducting cavity or LC resonator with Hamiltonian $\mathcal{H}_c=\Omega a^\dagger a$, being $\Omega$ the resonator frequency and $a$ the photon destruction operator.

The interaction between the spins and the photons is of Zeeman type and can be described by a generalized Dicke model~\cite{Generlized-Dicke,FLuis-ScalableArchitecture}, where the local quantized magnetic field generated by the supercurrents can be written as $\vec{B}_{\text{mw}}(\vec{r})=\vec{B}_{\text{rms}}(\vec{r})(a+a^\dagger)$, with $\vec{B}_{\text{rms}}(\vec{r})$ its zero-point fluctuations. The interaction Hamiltonian reads:
\begin{equation}
    \mathcal{H}_{I}\left(i\right)=\left(a^{\dagger}+a\right)\frac{\vec{\lambda}_{i}}{\sqrt{N}}\cdot\hat{g}\cdot\vec{S}_{i}
\end{equation}
where $\vec{\lambda}_{i}/\sqrt{N} = \mu_B \vec{B}_{\text{rms}}(\vec{r}_i)$.

In order to study the effective interactions between distant spins, mediated by the photons, we follow ref.~\cite{AGL-Spectroscopy} and write the total Hamiltonian:
\begin{equation}
    \mathcal{H}=\mathcal{H}_c+\sum_{i=1}^N \left[ \mathcal{H}_S(i)+\mathcal{H}_I(i)\right],\label{eq:Total-H}
\end{equation}
in the basis of Hubbard operators $X_i^{\vec{\alpha}}=|i,\alpha_1\rangle\langle i,\alpha_2|$, where $|i,\alpha\rangle$ is an eigenstate with energy $E_{i,\alpha}$ for the isolated $i$-th spin.

In this basis it is possible to derive an effective Hamiltonian using a Schrieffer-Wolff transformation~\cite{Wolff1966}, $\tilde{\mathcal{H}}=e^{\mathcal{S}}\mathcal{H}e^{-\mathcal{S}}$, which is valid in the dispersive regime and encodes the spin-photon interaction up to second order.

In contrast with the effective Hamiltonian from ref.~\cite{AGL-Spectroscopy}, here we include the presence of multiple spins and retain off-diagonal contributions as well. The former gives rise to their effective interaction mediated by the photons and the latter allows to calculate the dynamics for long time, which is when the off-diagonal terms become relevant.

The details of the lengthy, although straightforward derivation of the effective Hamiltonian using the Schrieffer-Wolff transformation are left for the~\hyperref[sec:Appendix-A]{Appendix~\ref*{sec:Appendix-A}}. 
The final expression for the effective Hamiltonian can be written as:
\begin{align}
    \tilde{\mathcal{H}}\simeq &\Omega a^{\dagger}a+\sum_{i=1}^{N}\sum_{\alpha=1}^{2S+1}E_{i,\alpha}X_{i}^{\alpha,\alpha}+\sum_{i=1}^{N}\sum_{\vec{\alpha}=1}^{2S+1}\delta E_{i,\vec{\alpha}}X_{i}^{\vec{\alpha}}\nonumber\\
    &+a^{\dagger}a\sum_{i=1}^{N}\sum_{\vec{\alpha}=1}^{2S+1}\delta\Omega_{i,\vec{\alpha}}X_{i}^{\vec{\alpha}}+\sum_{i,j\neq i}^{N}\sum_{\vec{\alpha},\vec{\beta}=1}^{2S+1}\tilde{J}_{i,j}^{\vec{\alpha},\vec{\beta}}X_{i}^{\vec{\beta}}X_{j}^{\vec{\alpha}}\nonumber\\
    &+\sum_{i=1}^{N}\sum_{\vec{\alpha}=1}^{2S+1}\left(\tilde{T}_{i,+}^{\vec{\alpha}}a^{\dagger}a^{\dagger}+\tilde{T}_{i,-}^{\vec{\alpha}}aa\right)X_{i}^{\vec{\alpha}},
    \label{eq:Effective-Hamiltonian}
\end{align}
Eq.~\eqref{eq:Effective-Hamiltonian} is a generalization of the effective Hamiltonian derived in Ref.~\cite{AGL-Spectroscopy}, now including off-diagonal corrections and the presence of several spins.
It must be mentioned that, for simplicity of the notation, the calculation of $\tilde{\mathcal{H}}$ assumes that all magnetic molecules have the same spin $S$, although their local environment can still be different. In contrast, if one is interested in the case of molecules with different spins, the limits in the sums just need to be changed to $2S_i+1$.

The first line in Eq.~\eqref{eq:Effective-Hamiltonian} contains all the free Hamiltonian terms, being the last one the shift in the unperturbed spin energy levels $E_{i,\alpha}$, produced by the virtual cavity photons (notice that it contains off-diagonal corrections that rotate the unperturbed eigenstates).
The second line contains the cavity-frequency shift produced by the state of the spins and the effective spin-spin interaction. 
The former term is crucial for the measurement of the state of the qudits using the cavity transmission, while the latter is critical to implement multi-qudit gates in hybrid c-QED architectures.
Finally, the last line encodes the role of two-photon processes, which can typically be neglected if the cavity occupation is small.

For the present purpose it is only relevant the explicit form of the effective spin-spin interaction ($E_{j,\vec{\alpha}}\equiv E_{j,\alpha_1}-E_{j,\alpha_2}$):
\begin{equation}
    \tilde{J}_{i,j}^{\vec{\alpha},\vec{\beta}}=\frac{\Omega\Lambda_{i}^{\vec{\beta}}\Lambda_{j}^{\vec{\alpha}}}{E_{j,\vec{\alpha}}^{2}-\Omega^{2}},
    \label{eq:Effective-Interaction}
\end{equation}
being $\Lambda_{i}^{\vec{\beta}}$ the spin-photon interaction projected onto the basis of Hubbard operators.
Concretely, its relation with the original parameters, in terms of $S_i^z$ eigenstates, is given in Eq.~\eqref{eq:Original-Parameters}.
As we are mainly interested in the effective interaction term, the explicit form of the other terms is provided in the~\hyperref[sec:Appendix-A]{Appendix~\ref*{sec:Appendix-A}}.

A new notation to simplify the study of the dynamics is now introduced. As we are interested in the dynamics of interacting spins, we rewrite the effective Hamiltonian in Eq.~\eqref{eq:Effective-Hamiltonian} as $\tilde{\mathcal{H}}=\tilde{\mathcal{H}}_0+\epsilon \tilde{\mathcal{V}}$. 
Here, $\tilde{\mathcal{H}}_0$ contains all the single spin terms, while $\tilde{\mathcal{V}}$ contains the effective spin-spin interaction ($\epsilon$ is just a free parameter that will help organize the perturbative series and will be taken to $1$ at the end of the calculations). 
This practical form can be easily obtained by tracing-out the photon sector in Eq.~\eqref{eq:Effective-Hamiltonian} and expressing the Hamiltonian in the basis of photon-dressed spin states, which include the energy shifts produced by the cavity photons, $\delta E_{j,\vec{\alpha}}$. 
Nevertheless, these shifts are not too relevant and tend to be small in the dispersive regime.


\section{Multiple-scales analysis for interacting qudits}
Multiple-scales analysis is a technique to study dynamical systems, which includes the renormalization of resonances~\cite{Multiple-scales-optics,PRR-bichromatic}. Its name comes from the fact that different orders in the expansion parameter correspond to different time-scales (ordered from the fastest to the slowest one). 
Importantly, the method of multiple-scales analysis can be applied to non-linear models as well~\cite{Multiple-scales-SpinBath}.

Here, we are interested in the calculation of the time-evolution operator $U(t)$ under the effective Hamiltonian $\tilde{\mathcal{H}}$.
Applying Multiple-scales analysis, we assume that for a small parameter $\epsilon$ and a Hamiltonian $\tilde{\mathcal{H}}=\tilde{\mathcal{H}}_0+\epsilon \tilde{\mathcal{V}}$, we can define a set of time-scales $\tau_n=\epsilon^n t$ and  expand the time-evolution operator in powers of the small parameter~\footnote{Notice that the physical small parameter when $\epsilon\to 1$ will be $\tilde{J}_{i,j}^{\vec{\alpha},\vec{\beta}}$}:
\begin{equation}
    U(t)=\sum_{n=0}^{\infty}\epsilon^n U_n(\vec{\tau})
\end{equation}
Inserting this expansion in the Schrödinger equation for the time-evolution operator $i\partial_t U(t)=\tilde{\mathcal{H}}U(t)$, and using the chain rule for the time-derivative, one finds the differential equation for the time-evolution operator at each order in $\epsilon$.
As for our purpose will be enough to consider linear corrections in $\epsilon$, we can easily check that the lowest and first order differential equations are:
\begin{align}
 i\partial_{\tau_{0}}U_{0}\left(\vec{\tau}\right)=&\tilde{\mathcal{H}}_{0}U_{0}\left(\vec{\tau}\right)\label{eq:Differential-equation1}\\
 i\partial_{\tau_{1}}U_{0}\left(\vec{\tau}\right)+i\partial_{\tau_{0}}U_{1}\left(\vec{\tau}\right)=&\tilde{\mathcal{H}}_{0}U_{1}\left(\vec{\tau}\right)+\tilde{\mathcal{V}}U_{0}\left(\vec{\tau}\right)\label{eq:Differential-equation2}
\end{align}
with $\vec{\tau}=(\tau_0,\tau_1)$.
The calculation of the time-evolution operator requires to solve these differential equations, and in the presence of secular terms (i.e., terms that grow unbounded with time), apply a renormalization procedure to encode their non-perturbative effect.

The solution to the unperturbed time-evolution operator $U_0(\vec{\tau})$ is given by:
\begin{equation}
    U_{0}\left(\vec{\tau}\right) = e^{-i\tilde{\mathcal{H}}_{0}\tau_{0}}u_{0}\left(\tau_{1}\right)\label{eq:Unperturbed0}
\end{equation}
The exponential describes the free evolution of each isolated spin, while the matrix $u_0(\tau_1)$ encodes the time evolution due to the slower time-scale $\tau_1$. 
This last term will be determined below from the renormalization procedure.

The first order correction is crucial to describe the implementation of multi-qudit gates. 
Its general expression is obtained from Eq.~\eqref{eq:Differential-equation2}, by defining $U_1(\vec{\tau})=e^{-i\tau_{0} \tilde{\mathcal{H}}_0}u_1(\vec{\tau})$ (this is a well-known trick to find the solution to inhomogeneous differential equations). 
The resulting differential equation for $u_1(\vec{\tau})$ is:
\begin{equation}
    \partial_{\tau_{0}}u_{1}\left(\vec{\tau}\right) = -ie^{i\tilde{\mathcal{H}}_{0}\tau_{0}}\tilde{\mathcal{V}}e^{-i\tilde{\mathcal{H}}_{0}\tau_{0}}u_{0}\left(\tau_{1}\right)-\partial_{\tau_{1}}u_{0}\left(\tau_{1}\right)\label{eq:perturbed0}
\end{equation}
where the matrix product $e^{i\tilde{\mathcal{H}}_{0}\tau_{0}}\tilde{\mathcal{V}}e^{-i\tilde{\mathcal{H}}_{0}\tau_{0}}$ can be interpreted as a transformation to a rotating frame.
Importantly, if some matrix elements are independent of $\tau_0$, the solution for $u_1(\vec{\tau})$ will display secular terms that grow unbounded with $\tau_0$.
This indicates that some transitions produced by the interaction are resonant and cannot be described perturbatively.
That is the reason why the slower time scale $\tau_1$ is introduced in multiple scales analysis, to encode the non-perturbative effect of resonances in the long time dynamics.
This can be easily seen in Eq.~\eqref{eq:perturbed0} from the fact that the last term produces a secular term of the form $\tau_0 \partial_{\tau_{1}}u_0(\tau_1)$.
Hence, if we choose $u_0(\tau_1)$ such that it cancels the secular terms from $e^{i\tilde{\mathcal{H}}_{0}\tau_{0}}\tilde{\mathcal{V}}e^{-i\tilde{\mathcal{H}}_{0}\tau_{0}}$, we can eliminate them from $U_1(\vec{\tau})$ by transferring their effect onto the unperturbed contribution $U_0(\vec{\tau})$, via $u_0(\tau_1)$.
Crucially, as the contribution $u_0(\tau_1)$ is non-divergent, this allows to extend the regime of validity of the solution to longer time scales.\\
Notice that the transformation to the interaction picture in Eq.~\eqref{eq:perturbed0}, and the following separation between secular and non-secular terms, indicates that the rotating wave approximation is contained within this method and that it naturally arises to first order. 
In addition, this method includes the role of counter-rotating terms in $U_1(\vec{\tau})$ and allows to systematically include higher order corrections, which can include additional resonances.\newline

Let us now particularize this method to the type of Hamiltonian under consideration. 
Starting from the effective Hamiltonian in Eq.~\eqref{eq:Effective-Hamiltonian}, we can trace-out the photon sector with a density matrix $\rho_p$ (typically a combination of the ground and first excited state in experimental setups).
Notice that the single spin contributions are cavity-dependent, because the splitting depends on its occupation, 
while the effective interaction does not.
Then, we can identify the unperturbed part of the effective Hamiltonian, $\tilde{\mathcal{H}}_{0}$, with the isolated spin terms:
\begin{equation}
    \tilde{\mathcal{H}}_{0} = \sum_{i=1}^{N}\sum_{\alpha=1}^{2S+1}\tilde{E}_{i,\alpha}\tilde{X}_{i}^{\alpha,\alpha}\label{eq:Hubbard2},
\end{equation}
where $\tilde{E}_{i,\alpha}$ is the energy for the $i$-th photon-dressed spin and $\tilde{X}_i^{\alpha,\alpha}$ the corresponding Hubbard operator.
Notice that this new Hubbard operators will be slightly rotated with respect to the initial ones, $X^{\alpha,\alpha}_i$, due to the presence of off-diagonal terms in Eq.~\eqref{eq:Effective-Hamiltonian}.

Finally, we can identify the effective interaction part of the Hamiltonian with the perturbation:
\begin{equation}
    \tilde{\mathcal{V}} = \sum_{i,j\neq i}^{N}\sum_{\vec{\alpha},\vec{\beta}=1}^{2S+1}\tilde{J}_{i,j}^{\vec{\alpha},\vec{\beta}}X_{i}^{\vec{\beta}}X_{j}^{\vec{\alpha}} = \sum_{i,j\neq i}^{N}\sum_{\vec{\alpha},\vec{\beta}=1}^{2S+1}\tilde{V}_{i,j}^{\vec{\alpha},\vec{\beta}}\tilde{X}_{i}^{\vec{\alpha}}\tilde{X}_{j}^{\vec{\beta}}
\end{equation}
where in the second equality we have expressed the interaction in the photon-dressed spin basis.\newline
With this identification, we can particularize the previous solutions to the system under discussion. 
The unperturbed time-evolution operator $U_0(\vec{\tau})$ results in:
\begin{equation}
    U_{0}\left(\vec{\tau}\right) = \prod_{i=1}^{N}\sum_{\alpha=1}^{2S+1}e^{-i\tilde{E}_{i,\alpha}\tau_{0}}\tilde{X}_{i}^{\alpha,\alpha}u_{0}\left(\tau_{1}\right),\label{eq:Unperturbed}
\end{equation}
As expected, the unperturbed solution describes the free evolution of each spin with phase factor $\tilde{E}_{i,\alpha}$.\newline
To write the first order solution $U_1(\vec{\tau})$, we start by calculating the product ($E_{j,\vec{\alpha}}\equiv E_{j,\alpha_1}-E_{j,\alpha_2}$):
\begin{equation}
    e^{i\tilde{\mathcal{H}}_{0}\tau_{0}}\tilde{\mathcal{V}}e^{-i\tilde{\mathcal{H}}_{0}\tau_{0}}=\sum_{i,i\neq j}^{N}\sum_{\vec{\mu},\vec{\nu}=1}^{2S+1}\tilde{V}_{i,j}^{\vec{\mu},\vec{\nu}}e^{i\left(\tilde{E}_{i,\vec{\mu}}+\tilde{E}_{j,\vec{\nu}}\right)\tau_{0}}\tilde{X}_{i}^{\vec{\mu}}\tilde{X}_{j}^{\vec{\nu}},\label{eq:phases}
\end{equation}
and determine the condition for the presence of secular terms. 
They will appear in Eq.~\eqref{eq:perturbed0} when the phase factor in Eq.~\eqref{eq:phases} cancels, which requires:
\begin{equation}
    \tilde{E}_{i,\vec{\mu}}+\tilde{E}_{j,\vec{\nu}}=0\label{eq:secular}
\end{equation}
This is the resonance condition for transitions between two qudit states.
Therefore, if a set of states fulfils the condition from Eq.~\eqref{eq:secular}, one needs to renormalize their contribution.
This can be done by imposing the following flow equation for the matrix $u_0(\tau_1)$:
\begin{equation}
    \dot{u}_{0}\left(\tau_{1}\right)=-i\sum_{i,j\neq i}^{N}\sum_{\langle \vec{\mu},\vec{\nu}\rangle=1}^{2S+1}\tilde{V}_{i,j}^{\vec{\mu},\vec{\nu}}\tilde{X}_{i}^{\vec{\mu}}\tilde{X}_{j}^{\vec{\nu}}u_{0}\left(\tau_{1}\right),\label{eq:flow1}
\end{equation}
where the summation over $\langle \vec{\mu},\vec{\nu} \rangle$ indicates that is restricted to states that fulfil Eq.~\eqref{eq:secular}.
Therefore, the lowest order time evolution operator, including the renormalization of resonances reads:
\begin{equation}
    U_{0}\left(\vec{\tau}\right) = e^{-i\tau_0\tilde{\mathcal{H}}_0} e^{-i\tau_1 \sum_{i,j\neq i}^{N}\sum_{\langle \vec{\mu},\vec{\nu}\rangle=1}^{2S+1}\tilde{V}_{i,j}^{\vec{\mu},\vec{\nu}}\tilde{X}_{i}^{\vec{\mu}}\tilde{X}_{j}^{\vec{\nu}}}\label{eq:Renormalized-Solution}
\end{equation}
In addition, the first order correction is given by:
\begin{align}
    U_{1}\left(\vec{\tau}\right)=\sum_{i,j\neq i}^{N}\sum_{\langle\langle \vec{\alpha},\vec{\beta} \rangle \rangle =1}^{2S+1}
    \tilde{V}_{i,j}^{\vec{\alpha},\vec{\beta}}(\tau_0)\tilde{X}_{i}^{\vec{\alpha}}\tilde{X}_{j}^{\vec{\beta}}u_0(\tau_1)\label{eq:time-evolution1}
\end{align}
where we have defined the time-dependent, non-secular contribution as:
\begin{equation}
    \tilde{V}_{i,j}^{\vec{\alpha},\vec{\beta}}(\tau_0) \equiv \tilde{V}_{i,j}^{\vec{\alpha},\vec{\beta}} \frac{e^{-i\left(\tilde{E}_{i,\alpha_{1}}+\tilde{E}_{j,\beta_{1}}\right)\tau_{0}}-e^{i\left(\tilde{E}_{i,\alpha_{2}}+\tilde{E}_{j,\beta_{2}}\right)\tau_{0}}}{\tilde{E}_{i,\vec{\alpha}}+\tilde{E}_{j,\vec{\beta}}}
\end{equation}
and the sum over $\langle\langle \vec{\alpha},\vec{\beta} \rangle \rangle$ is restricted to all the states that do not fulfil the resonance condition in  Eq.~\eqref{eq:secular}.

We can see that the free evolution for each spin qudit is renormalized by $u_0(\tau_1)$, which introduces a slower time-scale that completely dominates the dynamics at long times. 
The exponential form of this correction indicates that it is non-perturbative, while in contrast, as $U_1(\vec{\tau})$ is linear in $\tilde{V}_{i,j}^{\vec{\alpha},\vec{\beta}}$, its correction remains always small for arbitrary time.
For practical purposes this means that unwanted transitions will be kept under control over time, as far as they are not resonant.

This implies that, in order to capture the main features of the short-time and long-time dynamics, it is enough to use $U_0(\vec{\tau})$, because the short-time dynamics is dominated by the unperturbed part, while the long-time dynamics is controlled by the subset of states that are resonant. 
Also, it demonstrates that in general for qudits, tuning in and out of resonance the different qudit energy levels allows to switch on/off the multi-qudit gates.

In contrast with the qubits case, the multi-level structure of qudits also allows to control the available set of gates by just tuning the transitions that are in resonance.
For molecular spins this can be done by applying static or dynamic local magnetic fields, or by changing the orientation of the spin easy-axis with respect to the photon field.
In addition, the variety of magnetic molecules with different crystal fields also brings additional freedom to the set of gates that can be implemented.

Finally, notice that this approach also allows to perform reverse engineering. 
It would be possible to chose a particular multi-qudit gate, and then find most adequate alignment of the resonator with the easy axis of the molecule to implement the gate, or impose restrictions to its crystal field anisotropy, helping to select better molecules for quantum computation.

In the following, we discuss two examples that illustrate our results.
Also, to compare with the standard result for qubits, we analyze their case in full extent in the Appendices.  \hyperref[sec:Appendix-B]{Appendix~\ref*{sec:Appendix-B}} shows the derivation of their effective interaction and ~\hyperref[sec:Appendix-C]{Appendix~\ref*{sec:Appendix-C}} characterizes their dynamics using multiple-scales analysis.


\section{2-qutrit quantum gate}

We consider a toy model of molecular spins with $S=1$ (qutrits), quadratic longitudinal anisotropy and a transverse interaction with the photon field. The Hamiltonian reads:
\begin{equation}
    \mathcal{H}=\Omega a^\dagger a+\sum_{j=1}^N\left[D (S_j^z)^2+\Delta_{j}S_j^z+\xi_j (a^\dagger+a) S_j^x\right]\label{eq:S_1-initial-H},
\end{equation}
being $\Delta_j\equiv\mu_B g_z B_j^z$, $\xi_j=\lambda^x_j g_x$ and $E_{j,M_j}=D M_j^2+\Delta_j M_j$ with $M_j=\{\pm1,0\}$.
This situation is similar to the one encountered when NV-centers interact with cavity photons.
In that case, the quadratic longitudinal Stevens operator $D\simeq 2.87$GHz splits, in the ground state multiplet ($^{3}\text{A}_2$ state), the $m=\pm 1$ states from the $m=0$~\cite{NV-center}.

Although Eq.~\eqref{eq:Effective-Hamiltonian} and Eq.~\eqref{eq:Renormalized-Solution} are completely general for an arbitrary ensemble of qudits, for practical applications, one is mainly interested in quantum gates between specific pairs. 
This requires to switch-off the interactions with all the other spins in the resonator while simultaneously control the interaction for the pair of interest.
To decouple all the spins one just needs to take advantage of the result from Eq.~\eqref{eq:Renormalized-Solution} and set all the spins out of resonance using local magnetic fields.
\begin{figure}
    \centering
    \includegraphics[width=0.9\columnwidth]{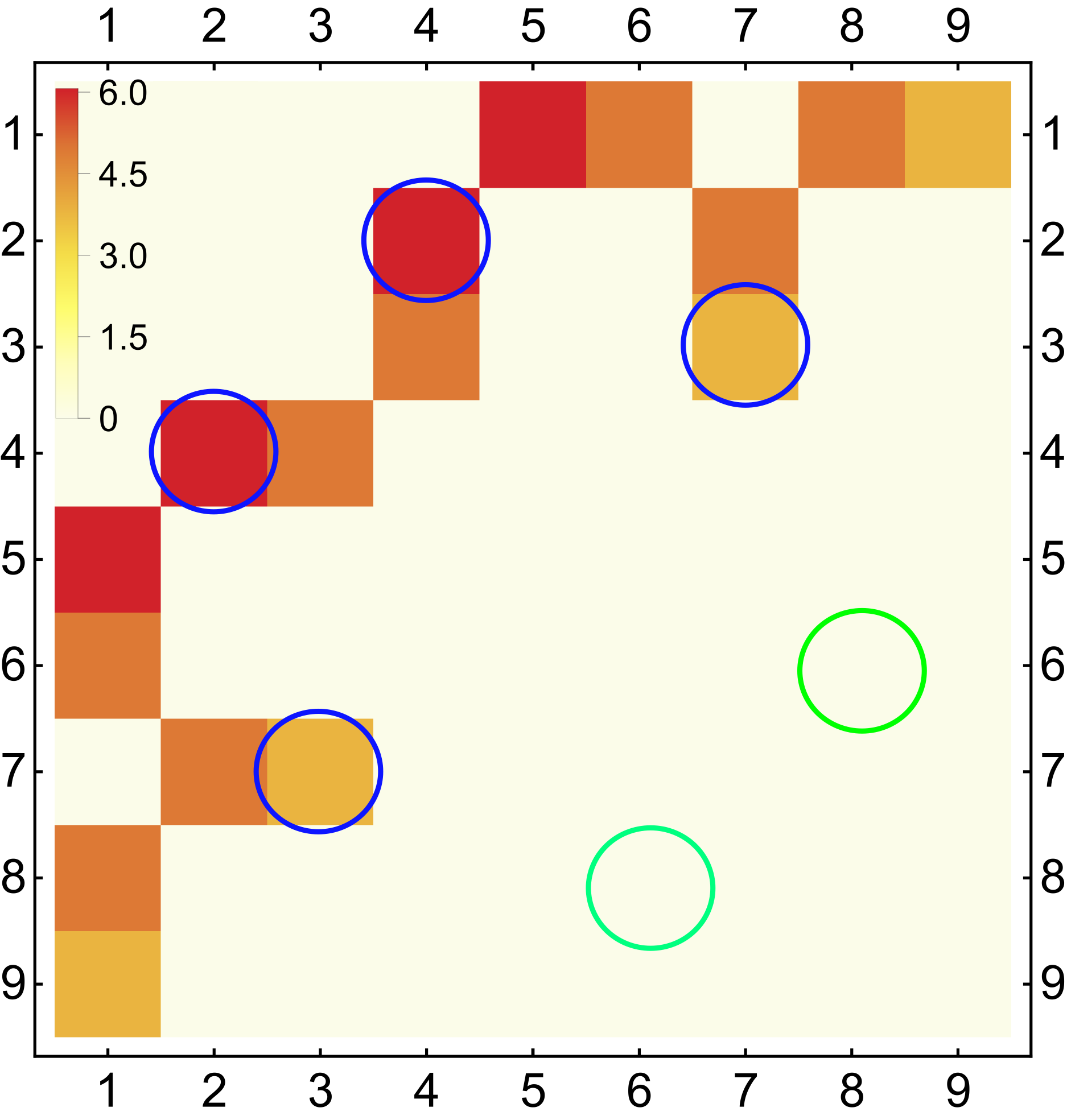}
    \caption{Components of the effective spin-spin interaction tensor, $\tilde{J}_{L,R}^{\mu,\nu}$, in kHz. The effective Ising type of interaction makes that only transitions that change their spin by $1$ can be coupled [that is why $a,b=\pm 1$ in Eq.~\eqref{eq:gate1}]. Blue circles indicate the transitions that fulfill the resonance condition and have non-zero coupling, while green circles indicate the ones that are resonant but not allowed by the symmetry of the effective spin interaction.}
    \label{fig:Interaction-S1}
\end{figure}

Now we focus on two particular spins that we label as left/right (L/R). Their effective interaction takes the following form [Cf Eq.~\eqref{eq:Effective-Interaction}]:
\begin{equation}
\tilde{\mathcal{V}}=\sum_{\mu,\nu=\pm}\tilde{J}_{L,R}^{\mu,\nu}\left(X_{L}^{0,\mu}+X_{L}^{\mu,0}\right)\left(X_{R}^{0,\nu}+X_{R}^{\nu,0}\right)
\end{equation}
with the coupling constant tensor given by (see Fig.~\ref{fig:Interaction-S1}):
\begin{equation}
    \tilde{J}_{L,R}^{\mu,\nu}=\frac{\xi_{L}\xi_{R}}{4}\left(\frac{\Omega}{E_{L,\mu}^{2}-\Omega^{2}}+\frac{\Omega}{E_{R,\nu}^{2}-\Omega^{2}}\right)
\end{equation}
Although this interaction changes strength depending on the qudit levels being involved, one can assume that for $\Omega\gg |E_{j,\alpha}|$, it approximately reduces to a transverse Ising interaction $J\simeq -\xi_{L}\xi_{R}/\Omega$. 
If in addition the energy shift produced by the cavity in each spin is negligible, the effective Hamiltonian from Eq.~\eqref{eq:Effective-Hamiltonian} becomes:
\begin{equation}
    \tilde{\mathcal{H}}\simeq \sum_{i=L,R}D\left(S_{i}^{z}\right)^{2}+\sum_{i=L,R}\Delta_{i} S_{i}^{z}+J S_{L}^{x}S_{R}^{x}\label{eq:Interaction-qutrit}
\end{equation}
This is a good approximation in the dispersive regime, where the original Hubbard operators for the spins used in Eq.~\eqref{eq:Effective-Hamiltonian} and the ones used in Eq.~\eqref{eq:Hubbard2} for the photon-dressed spins, are almost identical.
The full calculation can still be done for the exact interaction $J_{L,R}^{\pm,\pm}$ and eigenstates, but this is unnecessary for the current analysis.

From Eq.~\eqref{eq:Unperturbed} we can write the lowest order time-evolution operator as:
\begin{equation}
    U_{0}\left(\vec{\tau}\right)=\prod_{i=L,R}\sum_{M_i=-S}^{S}e^{-i\tau_{0}(D M_i^{2}+\Delta_{i}M_i)}X_{i}^{M_i,M_i}u_{0}\left(\tau_{1}\right)\label{eq:unperturbed1}
\end{equation}
The factor describes isolated spins oscillating with a phase that depends non-linearly on the quantum number $M_i$, while the second factor, $u_0 (\tau_1)$, corresponds to the non-perturbative correction, yet to be determined.

The calculation of the first order correction is straightforward and all its contributions are proportional to ($a,b=\pm 1$):
\begin{equation}
    e^{iH_{0}\tau_{0}}\tilde{\mathcal{V}}e^{-iH_{0}\tau_{0}}\propto X_{L}^{M_{L},M_{L}+a}X_{R}^{M_{R},M_{R}+b} e^{i\Phi\tau_0}\label{eq:gate1}
\end{equation}
The set of available transition operators for this model is a direct consequence of the transverse Ising-like interaction in $\tilde{\mathcal{H}}$ and can be modified by controlling the original spin-photon interaction.

We now focus on the phase factor multiplying each interaction term, which is given by:
\begin{equation}
    \Phi=E_{L}\left(M_{L}\right)+E_{R}\left(M_{R}\right)-E_{L}\left(M_{L}+a\right)-E_{R}\left(M_{R}+b\right),
\end{equation}
and as previously discussed in Eq.~\eqref{eq:secular}, controls the appearance of secular terms. 
In this particular case, the condition for a resonance becomes:
\begin{equation}
   \left(2DM_{L}+\Delta_{L}\right)a+\left(2DM_{R}+\Delta_{R}\right)b+2D=0\label{eq:qutrit-secular}
\end{equation}
The presence of the quantum numbers $M_j$ in this condition indicates that the spectrum is unequally spaced and that the resonance condition is different for each pair of levels.
In particular in this case we have the following types of operators that can produce resonant interactions (we use below that because $b=\pm1$, we can write $b=1/b$):
\begin{itemize}
    \item For $a=b$ : $X_{L}^{M_{L},M_{L}+b}X_{R}^{M_{R},M_{R}+b}$ with the resonance condition $M_{L}+M_{R}=-\frac{\Delta_{L}+\Delta_{R}}{2D}-b$.
    \item For $a=-b$ : $X_{L}^{M_{L},M_{L}-b}X_{R}^{M_{R},M_{R}+b}$ with the resonance condition $M_{L}-M_{R}=-\frac{\Delta_{L}-\Delta_{R}}{2D}+b$.
\end{itemize}
As $b=\pm1$ and the quantum numbers $M_j$ can only take discrete values, the resonances will only happen at specific values of $\Delta_L\pm\Delta_R$ which are multiples of the longitudinal anisotropy $2D$.
Furthermore, this value will also determine the relation between $M_L$ and $M_R$, and control the final form of the resonant interaction operators.

Notice that this is unimportant for the qubits case, or more generally, for multi-level systems with equally spaced levels.
To see this, notice that for $D=0$ the resonance condition reduces to $\Delta_{L}a=-\Delta_{R}b$, which does not involve $M_j$ and therefore leads to many simultaneous resonant transitions of the form $X_{L}^{M_{L},M_{L}+a}X_{R}^{M_{R},M_{R}+b}$, with arbitrary values of $M_j$.
This is not very advantageous for practical applications, as typically one is interested in addressing specific transitions.\newline
In contrast, one advantage of molecular spins with crystal field anisotropy is that their non-linear terms impose additional constraints between $M_L$ and $M_R$, reducing the set of levels that fulfil the resonance condition.
This allows to address specific transitions by tuning the energy levels with local fields.
In addition, Eq.~\eqref{eq:gate1} can be used to identify the full set of gates that can be implemented between pairs of qudits for a given effective interaction and crystal field anisotropy.

\begin{figure}
    \centering
    \includegraphics[width=1\columnwidth]{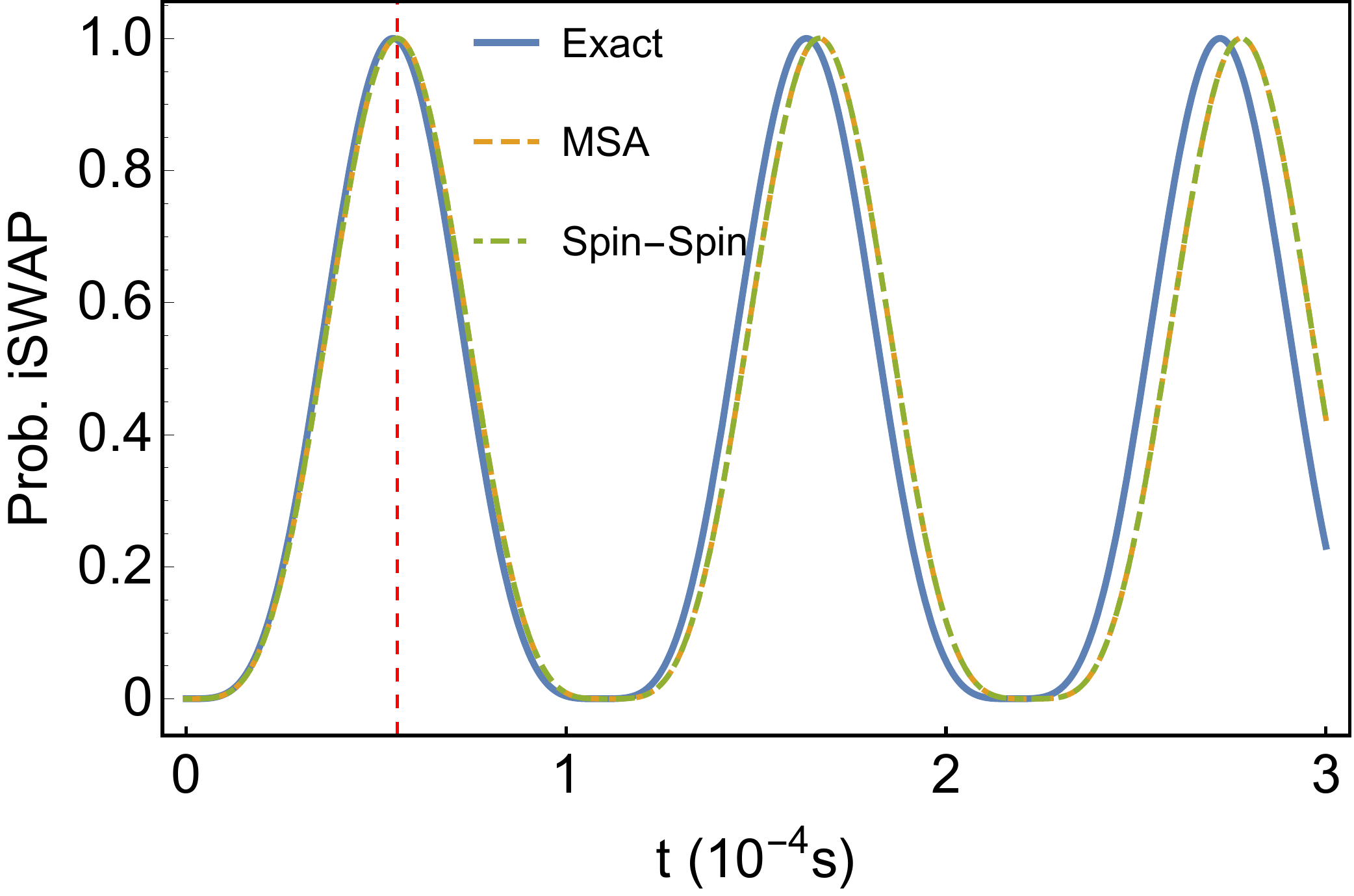}
    \caption{Probability for an iSWAP operation $|\downarrow,0 \rangle \to |0,\downarrow \rangle$ between a pair of NV-centers with $\Delta_j=0.007$T, $\xi_j=0.01$GHz and $\Omega=5$GHz. The solid line corresponds to the exact calculation using Eq.~\eqref{eq:S_1-initial-H}. The dot-dashed line corresponds to the exact time evolution using Eq.~\eqref{eq:Interaction-qutrit}, which completely agrees with the non-perturbative solution $U_0$(t) from Eq.~\eqref{eq:unperturbed1}. The vertical dashed line shows the estimated $\tau_{\text{gate}}=\pi/J$.}
    \label{fig:SWAP-S_1}
\end{figure}

Now we explicitly calculate the non-perturbative correction due to the secular terms, $u_0(\tau_1)$, and determine the time required for the implementation of the corresponding 2-qudit gate. 
We focus on the symmetric field configuration $\Delta_j=\Delta$, which leads to a SWAP-type of interaction $X_{L}^{M\pm1,M}X_{R}^{M,M\pm1}$ for the terms with $a=-b$.
In contrast, the terms with $a=b$ additionally require $r=\Delta/D$ to be an integer, in which case they take the form $X_{L}^{-M_{R}-r\mp1,-M_{R}-r}X_{R}^{M_{R},M_{R}\pm1}$.\newline
Therefore, if we take $\Delta/D\notin\mathbb{Z}$, we can make a SWAP-type of interaction, and the non-perturbative correction to the unperturbed time-evolution operator is given by Eq.~\eqref{eq:unperturbed1} with:
\begin{equation}
    u_0(\tau_1)=e^{-i\tau_1\frac{J}{2}\sum_{\nu=\pm}\left(X_{L}^{0,\nu}X_{R}^{\nu,0}+X_{L}^{\nu,0}X_{R}^{0,\nu}\right)}\label{eq:iSwap}
\end{equation}
If one re-writes Eq.~\eqref{eq:iSwap} in the basis $|M_L,M_R\rangle$, it can be seen that implements the following gate in both, the subspace of $M_j=\left\{0,\uparrow\right\}$ and the subspace of $M_j=\left\{0,\downarrow\right\}$:
\begin{equation}
    \left(\begin{array}{cccc}
1 & 0 & 0 & 0\\
0 & \cos\left(\frac{J\tau_{1}}{2}\right) & -i\sin\left(\frac{J\tau_{1}}{2}\right) & 0\\
0 & -i\sin\left(\frac{J\tau_{1}}{2}\right) & \cos\left(\frac{J\tau_{1}}{2}\right) & 0\\
0 & 0 & 0 & 1
\end{array}\right)
\end{equation}
Therefore, if the condition for the resonance $\Delta_L=\Delta_R$ is kept during a time $\tau_{\text{gate}}\sim \pi/J$, it applies an iSWAP gate to a subset of levels of the pair of qudits (while this subset is selected by fixing the static field correctly). 
This can be seen in Fig.~\ref{fig:SWAP-S_1}, where we have calculated the probability to perform an iSWAP operation: $|\downarrow,0 \rangle \to |0,\downarrow \rangle$. The non-perturbative result from multiple-scales analysis (dashed) provides an excellent agreement with the exact result (solid).\newline
During this time, the states $|\uparrow,\downarrow\rangle$ and $|\downarrow,\uparrow\rangle$ remain invariant because they are not coupled by the effective interaction tensor.
Importantly, the time required to perform the operation scales inversely with the square of the original spin-photon interaction $\lambda_j^x$.
As this interaction is typically small (of the order of a few Hz), that is why decoherence must be largely suppressed in order to implement the gates in experimental setups.
In summary, for this simple example we have shown that in qutrits it is possible to implement several gates. In particular, we have analyzed the case with symmetrical splitting $\Delta_{j}$, which is important for experimental setups, because it can produce an iSWAP gate.
It shows that three requirements must be simultaneously accounted for: 
\begin{itemize}
    \item The states of interest must fulfill the resonance condition from Eq.~\eqref{eq:secular}.
    \item The alignment of the easy axis of the molecule and the photon field must produce non-vanishing coupling, $\tilde{J}_{L,R}^{\mu,\nu}\neq 0$, between the resonant states.
    \item The time required for the relevant resonance must be shorter that $T_2$.
\end{itemize}

In addition, notice that with qutrits is possible to consider a different resonance condition than $\Delta_L=\Delta_R$ and obtain a different 2-qudit gate.
\section{Logical gates in $\text{GdW}_{30}$}
We now consider the experimentally motivated case of two lanthanide single-ion magnets of $\text{GdW}_{30}$~\cite{GdW-Molecule} polyoxometalate clusters coupled to the same photonic cavity. These molecules have $S=7/2$ and both, in-plane and longitudinal anisotropy, described by the Stevens operators $O_{2}^{0}=3\left(S^{z}\right)^{2}-S\left(S+1\right)$ and $O_{2}^{2}=\left(S^{x}\right)^{2}-\left(S^{y}\right)^{2}$. 
The Hamiltonian for an isolated molecular spin reads:
\begin{equation}
\mathcal{H}_{S}=\frac{D_{1}}{3}O_{2}^{0}+E_{2}O_{2}^{2}-g\mu_{B}\vec{B}\cdot\vec{S}
\end{equation}
The values of the different parameters have been experimentally determined to $D_{1}=1.281\text{GHz}$, $E_{2}=0.294\text{GHz}$ and $g=2$. 
\begin{figure}
\centering
\includegraphics[width=1\columnwidth]{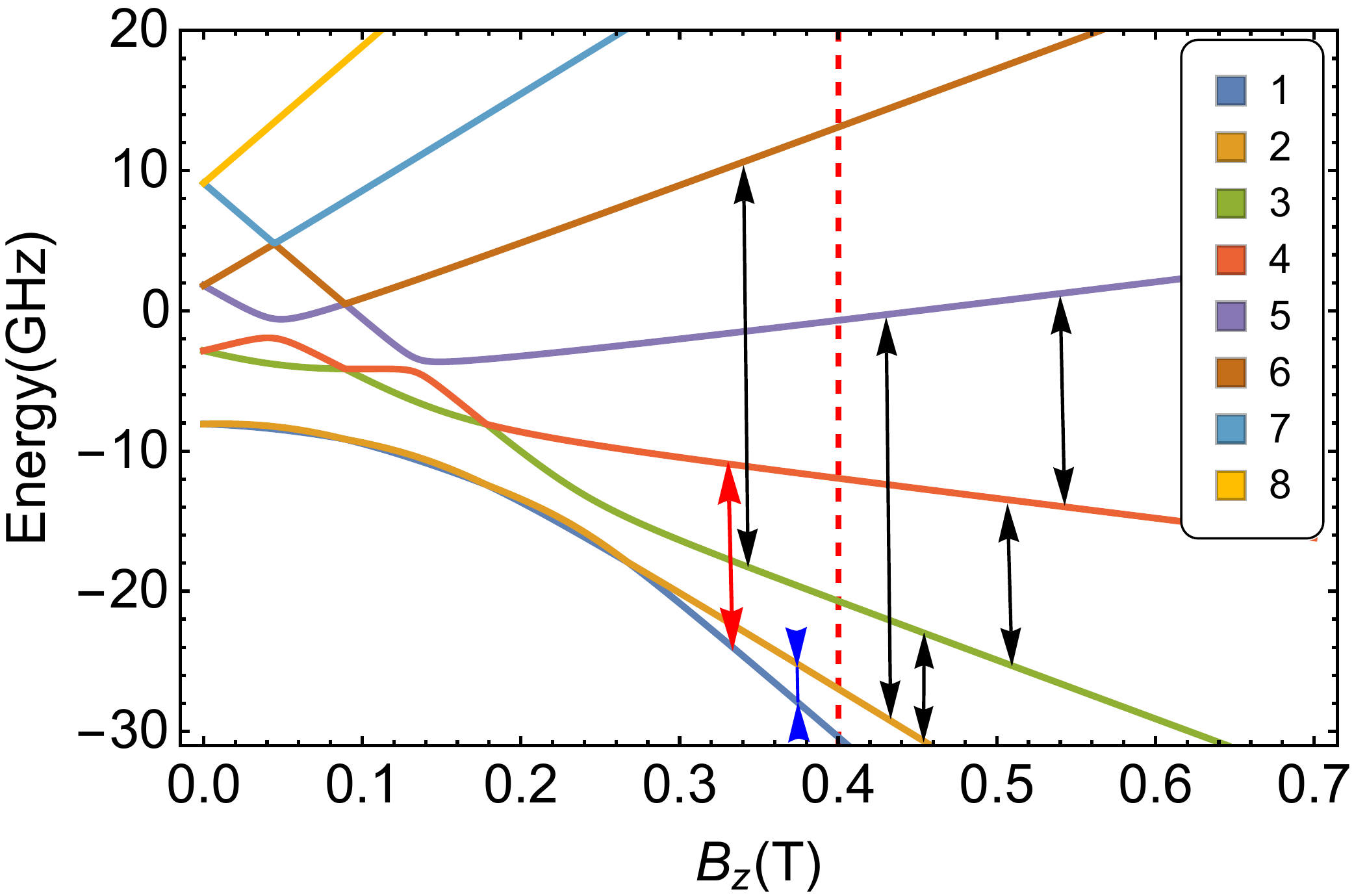}
\caption{\label{fig:Fig1}Level structure of $\textrm{GdW}_{30}$ vs $B_z$. The 2S+1 levels are labeled in order from the ground $(1)$ to the highest excited state $(8)$. The vertical red-dashed line indicates the field chosen below to perform the 2-qudit operations. The arrows indicate some of the resonant transitions between pairs of qudits at $B_z=0.4$T. The red/blue arrows indicate the 2-qudit transitions chosen to simulate the dynamics in Fig.~\ref{fig:Fig3} 
($|1,4\rangle \leftrightarrow |4,1\rangle$ and $|1,2\rangle \leftrightarrow |2,1\rangle$, respectively).}
\end{figure}
The energy level structure of an isolated $\textrm{GdW}_{30}$ molecular spin is shown in Fig.~\ref{fig:Fig1}, as a function of the longitudinal field. 
It can be seen how the crystal field makes the energy levels unequally spaced and their non-linear dependence on $B_z$.
This allows to tune the energy levels of each molecule, and in consequence, the implementation of different operations between the qudits, with the distinctive feature that the non-linear dependence can highly modify their properties.

In this case, when the two molecules are coupled through the resonator, the effective interaction from Eq.~\eqref{eq:Effective-Interaction} is characterized by the spin-photon interaction tensor $\Lambda_i^{\vec\alpha}$ and the cavity frequency $\Omega$. 
We consider a frequency of $\Omega=3$GHz, which is smaller than all the available transitions at $B_z=0.4$T and makes our description in the dispersive regime correct. 
\begin{figure}
\includegraphics[width=1\columnwidth]{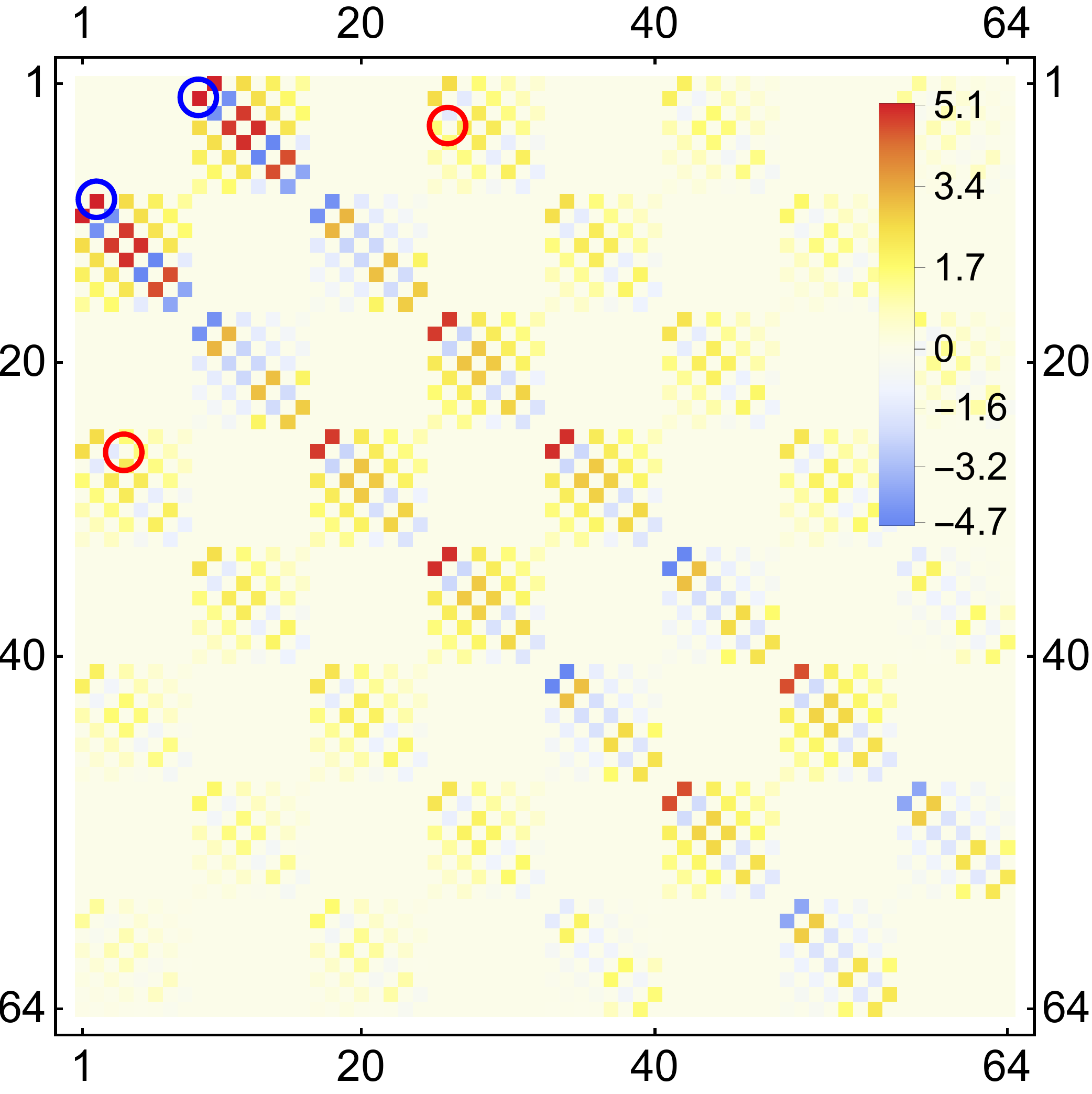}
\caption{\label{fig:Fig2}Tensor components of the effective interaction $J_{L,R}^{\vec{\alpha},\vec{\beta}}$ ($10^{-4}$GHz) for $B_z=0.4$T, $\Omega=3$GHz and $\vec{\lambda}_i=(10^{-2},0,0)$GHz. The basis is such that the first column elements vary $\alpha_1$ and $\beta_1$ from $1$ to $8$ while keeping the others fixed, and the first row elements vary $\alpha_2$ and $\beta_2$ from $1$ to $8$. Circles in red/blue correspond to the selected transitions $|1,4\rangle \leftrightarrow |4,1\rangle$ and $|1,2\rangle \leftrightarrow |2,1\rangle$, respectively.}
\end{figure}
The resulting effective interaction tensor is shown in Fig.~\ref{fig:Fig2}. It shows that the effective interaction connects many levels, although their coupling strength is generally small. This is because their interaction is proportional to $|\vec{\lambda}_i|^2$. Also, certain transitions remain decoupled due to the particular form of the crystal field anisotropy and the transverse spin-photon coupling.

Presently, the spin-photon coupling for a single molecular spin is in the range between Hz and kHz~\cite{FLuis-MolecularSpins&QC,FLuis-ScalableArchitecture}.
Although this results in values of $J_{L,R}^{\vec{\alpha},\vec{\beta}}$ smaller than the photon cavity loss (which is of the order of kHz), future prospects indicate that it will be possible to reach larger values to operate in the dispersive regime, for example, by combining electronic and nuclear spins, and nano-constrictions~\cite{Constriction,HighCooperativity}. 
In particular, if $\vec{\lambda}_i$ is pushed to the MHz range, the interaction will be of the order of $J_{L,R}^{\vec{\alpha},\vec{\beta}}\sim 10^{2}$kHz, which is larger than the typical photon loss and provides a time-window to act on the qudits. 
This is the value that we considered in Fig.~\ref{fig:Fig2} and that will be used from now on.

Once the system has entered in the dispersive regime, the small value of the couplings in Fig.~\ref{fig:Fig2} is not critical. 
As shown above, only resonant interactions can produce 2-qudit gates, and their coupling strength just affects the time required for their implementation.
Hence, from Fig.~\ref{fig:Fig2} one needs to isolate the resonant interactions [i.e., the ones that fulfil Eq.~\eqref{eq:secular}] and solve the flow equation to obtain the non-perturbative correction $u_0(\tau_1)$ from Eq.~\eqref{eq:flow1}.

\begin{figure}
\centering
\includegraphics[width=1\columnwidth]{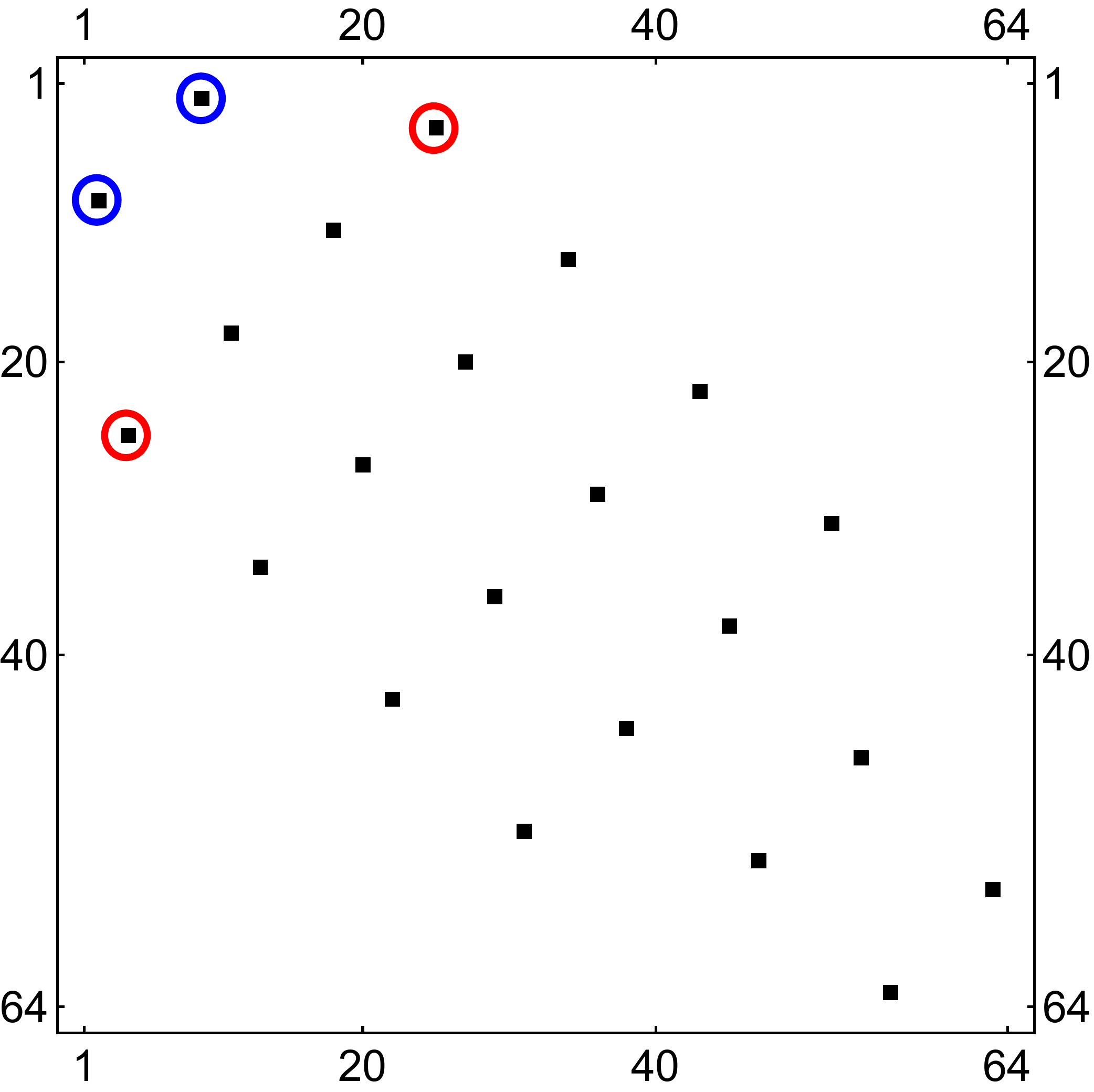}
\caption{\label{fig:Fig3}Resonant transitions between the two $\textrm{GdW}_30$ molecules for $H_z=0.4$T, $\Omega=3$GHz and $\lambda_i^x=10^{-2}$GHz. We choose the transitions $|1,4\rangle \leftrightarrow |4,1\rangle$ and $|1,2\rangle \leftrightarrow |2,1\rangle$ (surrounded by red and blue circles, respectively) to study their time-dependence. Other choices of $B_z$ would lead to different resonant transitions and then, to different quantum gates.}
\end{figure}

For our choice with an identical longitudinal field of $B_z=0.4$T in both molecules, as indicated by the vertical dashed line in Fig.~\ref{fig:Fig1}, we find that $11$ transitions are resonant.
Notice that even in this highly symmetric condition with two identical molecules, not all transitions fulfilling the resonance condition from Eq.~\eqref{eq:secular} contribute, due to the interplay between the crystal field anisotropy and the alignment with the photon field.
This is shown in Fig.~\ref{fig:Fig3}, where the black squares correspond to the resonant terms extracted from Fig.~\ref{fig:Fig2}.
These resonant interactions are of SWAP-type and some of them are illustrated with arrows in Fig.~\ref{fig:Fig1}.

Notice that the subset of resonant interactions has a wide variety of values, indicating the presence of very different time-scales for the resulting quantum gates.
To show this, we choose two cases of resonant interactions with very different interaction strength $J_{L,R}^{\vec{\alpha},\vec{\beta}}$.
The blue circles in Fig.~\ref{fig:Fig2} indicate a transition that swaps the states $|1\rangle$ and $|2\rangle$ of the two spins. In contrast, the red circles indicate a swap transition between states $|1\rangle$ and $|4\rangle$ which is also resonant, but has smaller coupling strength (cf with color code in Fig.~\ref{fig:Fig2}). These two transitions are also marked in Fig.~\ref{fig:Fig3} to show that they are resonant.

It is important to stress that one could naively think that the interaction marked with a red circle in Fig.~\ref{fig:Fig2} is negligible due to its small strength, if compared with others with a larger value.
This would be the case for short time evolution only. Beyond that, the smaller but resonant transition would take over.
That is why it is important to use a non-perturbative approach  to describe the dynamics, such as multiple scales analysis.

Concretely, what the strength of the interaction for the resonant terms characterizes is the time required to implement the corresponding gate. This means that the SWAP gate $|1,4\rangle \leftrightarrow |4,1\rangle$ requires longer time than the SWAP gate $|1,2\rangle \leftrightarrow |2,1\rangle$, although both can be implemented. 

Obviously it is crucial to compare this gate-implementation estimated time with the $T_2$ of the spins and the photon losses of the cavity.
For example, it will allow to decide what gates are experimentally feasible for a specific molecular spin at a fixed $\vec{B}_j$ configuration.

\begin{figure}
\centering
\includegraphics[width=1\columnwidth]{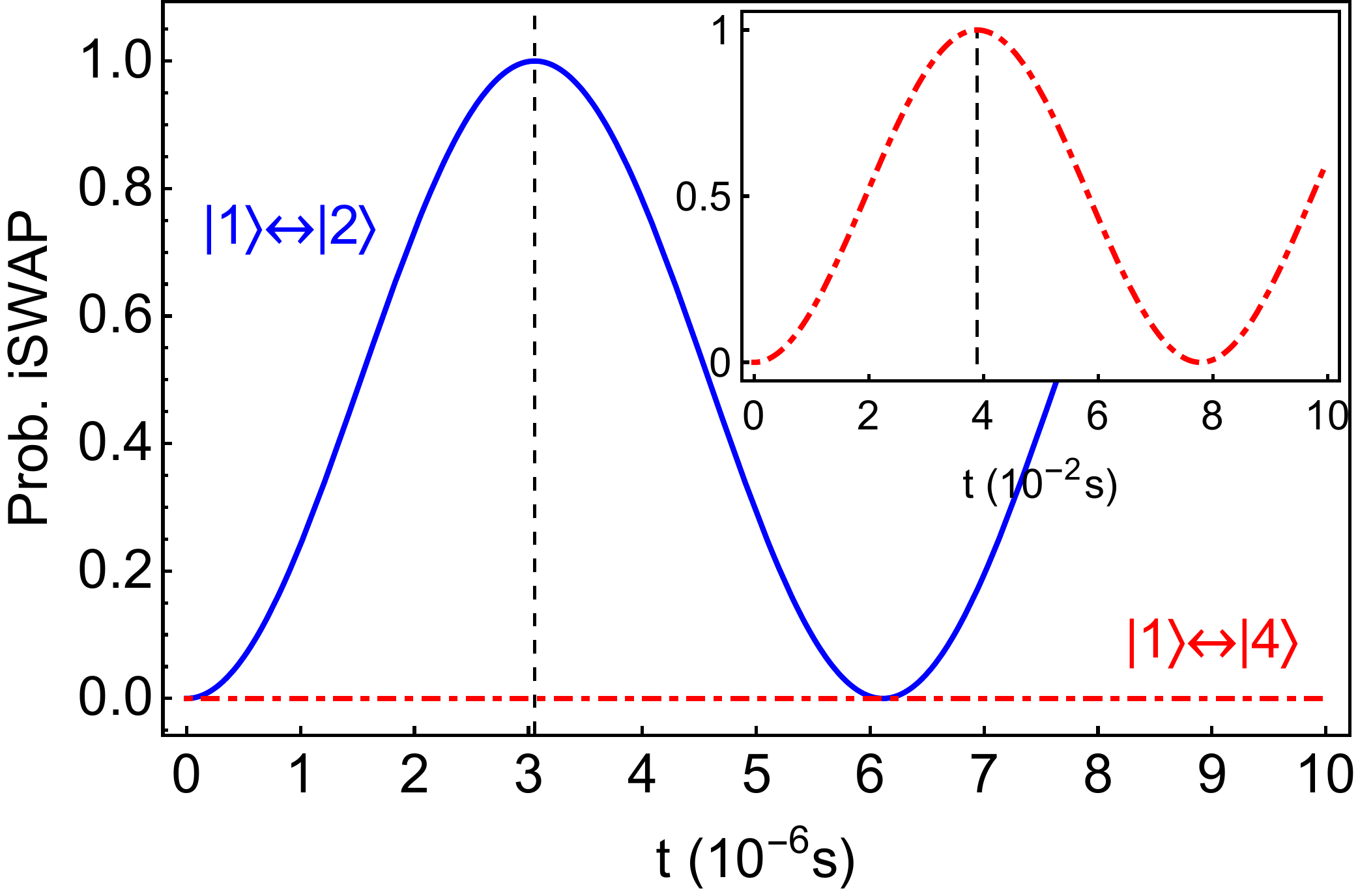}
\caption{\label{fig:Fig4}Probability to perform the iSWAP operation $|1,2\rangle \to |2,1\rangle$ (solid blue) from $U_0(t)$. The vertical dashed line indicates the estimated time from the inverse of the effective interaction.
The inset shows the probability for the iSWAP operation $|1,4\rangle \to |4,1\rangle$ (dashed red), which due to its smaller coupling strength, requires a longer time (see Fig.~\ref{fig:Fig2}). Then, experimentally is more feasible to implement the iSWAP gate between states $|1,2\rangle \leftrightarrow |2,1\rangle$, but this can be controlled with the local fields $\vec{B}_i$.}
\end{figure}

To confirm our predictions, we calculate the time evolution for the non-perturbative correction $U_0(t)$ [see Eq.~\eqref{eq:Renormalized-Solution}] and study the two resonant transitions selected in Fig.~\ref{fig:Fig3}.
This is shown in Fig.~\ref{fig:Fig4}, where we plot the probability for the corresponding SWAP operation. 
We find that in both cases the correction is non-perturbative, as it oscillates between $0$ and $1$. 
Furthermore, analyzing its real and complex parts we can see that it exactly corresponds to an iSWAP operation.
The first maximum indicates the minimum time $t_{\text{iSWAP}}\sim \pi/2J_{L,R}^{\vec{\alpha},\vec{\beta}}$ required to perform the iSWAP operation (indicated by a vertical dashed line). 
The operation $|1,2\rangle \leftrightarrow |2,1\rangle$ (blue) requires a time of the order of $\mu$s, while the operation $|1,4\rangle \leftrightarrow |4,1\rangle$ (inset, dashed-red) requires of the order of $0.1$s.
Therefore, it would only be experimentally feasible to implement the iSWAP between $|1,2\rangle \leftrightarrow |2,1\rangle$ in the $\text{GdW}_{30}$ for the chosen parameters of cavity frequency and field configuration.

\section{Conclusions:}
We have obtained a general form for the effective Hamiltonian describing a set of qudits interacting via a single mode cavity in the dispersive regime. 
We have particularized our analysis to the case of molecular spins with arbitrary $S$, where non-linear contributions due to crystal anisotropy are also important. 
From this result we have determined the effective qudit-qudit interaction in terms of the microscopic parameters of the system and studied the implementation of quantum gates from the time evolution of the system. 
We find that qudits largely enhance the computational possibilities of the setup, with respect to the qubit case, and the non-linearities from crystal field anisotropy can help to design quantum gates between specific sets of levels. In addition, our study of the dynamics in terms of multiple-scales analysis can be used in a wide number of cases not typically covered by the literature, such as for asymmetric splitting configurations or for different molecules. 
Finally, we have considered in detail the case of a pair of NV-centers and $\text{GdW}_{30}$ molecules, widely used in quantum technologies for their interesting properties, and we have shown that tuning the longitudinal field we can select various sets of transitions to implement different gates.
Crucially, our analysis allows us to study their quantum dynamics in detail and extract the time required for the implementation of the quantum gates in terms of the microscopic parameters. This will help for an efficient design of the quantum architectures and to determine the decoherence threshold required for their practical use.
Importantly, these results are also valid in other architectures comprising qudits with unequally spaced levels in the dispersive regime.
\begin{acknowledgments}
We thank D. Zueco, F. Luis and G. F. Peñas for insightful discussions.
We acknowledge the funding from the Spanish MICIN grant PCI2018-093116 (MCIU/AEI/FEDER, UE) and the  CSIC Interdisciplinary Thematic Platform (PTI+) on Quantum Technologies (PTI-QTEP+).
\end{acknowledgments}
\begin{widetext}
\appendix
\section{Derivation of the effective Hamiltonian:}\label{sec:Appendix-A}
We apply a Schrieffer-Wolff(S-W) transformation $\mathcal{S}$ to derive an effective Hamiltonian encoding, up to second order, the interaction with the cavity photons:
\begin{equation}
    \tilde{\mathcal{H}}=e^{\mathcal{S}}\mathcal{H}e^{-\mathcal{S}} \simeq \mathcal{H}_0+\frac{1}{2}\left[\mathcal{S},\mathcal{V}\right]
    \label{eq:Schrieffer-Wolff-Appendix}
\end{equation}
where $\mathcal{V}=\sum_{i=1}^{N}\mathcal{H}_I\left(i\right)$ corresponds to the interaction term and $\mathcal{H}_0=\sum_{i=1}^{N}\mathcal{H}_S\left(i\right)+\mathcal{H}_c$ contains the Hamiltonian for the isolated molecules and the resonator.
To obtain Eq.~\eqref{eq:Schrieffer-Wolff-Appendix} one needs to impose the condition $\left[\mathcal{S},\mathcal{H}_0\right]=-\mathcal{V}$ and assume weak spin-photon interaction to truncate the higher order terms. This will allow to fix the free parameters in the ansatz for $\mathcal{S}$.

As previously mentioned in the main text, in order to simplify the treatment of the non-linear spin terms produced by the Steven operators, it is useful to work in the Hubbard operators basis for the isolated molecules.
Hence, the Hamiltonian for an isolated molecule simply is:
\begin{equation}
\mathcal{H}_S\left(i\right)=\sum_{\alpha=-S}^{S}E_{i,\alpha}X_{i}^{\alpha,\alpha}\label{eq:Hamiltonian-S}
\end{equation}
where $X_i^{\vec{\alpha}}=|i,\alpha_1\rangle\langle i,\alpha_2|$ is the transition operator from eigenstate $|\alpha_2\rangle$ to $|\alpha_1\rangle$ in the $i$-th spin.
Analogously, we write the interaction operator in the Hubbard operators basis:
\begin{equation}
    \mathcal{H}_{I}\left(i\right)=\left(a^{\dagger}+a\right)\sum_{\vec{\alpha}=1}^{2S+1}\Lambda_{i}^{\vec{\alpha}}X_{i}^{\vec{\alpha}}
\end{equation}
The relation with the original parameters, expressed in terms of $S_i^z$ eigenstates $|S_i,M_i\rangle$, is given by:
\begin{eqnarray}
\Lambda_{i}^{\vec{\alpha}} & = & \frac{\lambda_{i}^{z}g_{z}}{\sqrt{N}}\sum_{M_i=-S_i}^{S_i}M_{i}c_{\alpha_{1},M_i}c_{\alpha_{2},M_i}^{\ast}\nonumber\\
 &  & +\sum_{M_i=-S_i}^{S_i}\gamma_{S_i,M_i}\frac{\lambda_{i}^{x}g_{x}-i\lambda_{i}^{y}g_{y}}{2\sqrt{N}}c_{\alpha_{1},M_i+1}c_{\alpha_{2},M_i}^{\ast}\nonumber \\
 &  & +\sum_{M_i=-S_i}^{S_i}\gamma_{S_i,M_i}\frac{\lambda_{i}^{x}g_{x}+i\lambda_{i}^{y}g_{y}}{2\sqrt{N}}c_{\alpha_{1},M_i}c_{\alpha_{2},M_{i}+1}^{\ast}\label{eq:Original-Parameters}
\end{eqnarray}
being $\gamma_{S_i,M_i}=\sqrt{S_i\left(S_i+1\right)-M_i\left(M_i+1\right)}$, $c_{\alpha,M_i}=\langle i,\alpha|S_i,M_i\rangle$ and $|S_i,M_i\rangle$ the $S_{i}^{z}$ eigenstates (do not confuse $\gamma_{S_i,M_i}$ with the cavity losses $\gamma$).
It will be shown below that the Hubbard basis is very useful to describe the dynamics as well.

Now one can fix the ansatz for the transformation matrix $\mathcal{S}$ to:
\begin{equation}
\mathcal{S}=\sum_{i=1}^{N}\sum_{\vec{\beta}=1}^{2S_i+1}\left(\frac{\Lambda_{i}^{\vec{\beta}}}{E_{i,\vec{\beta}}+\Omega}a^{\dagger}+\frac{\Lambda_{i}^{\vec{\beta}}}{E_{i,\vec{\beta}}-\Omega}a\right)X_{i}^{\vec{\beta}} \label{eq:S-W}
\end{equation}
where $E_{i,\vec{\beta}}=E_{i;\beta_1,\beta_2} \equiv E_{i,\beta_1}-E_{i,\beta_2}$. 
The effective Hamiltonian from Eq.~\eqref{eq:Schrieffer-Wolff-Appendix}, is valid in the dispersive regime (i.e., for $\gamma \ll \Lambda_{i}^{\vec{\alpha}} \ll \left|\left|E_{i,\vec{\alpha}}\right|-\Omega\right|$), being $\gamma$ the cavity loss,  and takes the following form:
\begin{align}
    \tilde{\mathcal{H}}\simeq &\Omega a^{\dagger}a+\sum_{i=1}^{N}\sum_{\alpha=1}^{2S+1}E_{i,\alpha}X_{i}^{\alpha,\alpha}+\sum_{i=1}^{N}\sum_{\vec{\alpha}=1}^{2S+1}\delta E_{i,\vec{\alpha}}X_{i}^{\vec{\alpha}}\nonumber\\
    &+a^{\dagger}a\sum_{i=1}^{N}\sum_{\vec{\alpha}=1}^{2S+1}\delta\Omega_{i,\vec{\alpha}}X_{i}^{\vec{\alpha}}+\sum_{i,j\neq i}^{N}\sum_{\vec{\alpha},\vec{\beta}=1}^{2S+1}\tilde{J}_{i,j}^{\vec{\alpha},\vec{\beta}}X_{i}^{\vec{\beta}}X_{j}^{\vec{\alpha}}\nonumber\\
    &+\sum_{i=1}^{N}\sum_{\vec{\alpha}=1}^{2S+1}\left(\tilde{T}_{i,+}^{\vec{\alpha}}a^{\dagger}a^{\dagger}+\tilde{T}_{i,-}^{\vec{\alpha}}aa\right)X_{i}^{\vec{\alpha}}
    \label{eq:Effective-Hamiltonian-Appendix}
\end{align}
where we have assumed that all molecules have the same spin $S$, to simplify the notation. Otherwise, the upper limits in the sums must be changed accordingly to $2S_i+1$.

Eq.~\eqref{eq:Effective-Hamiltonian} is a generalization to the one derived in ref.~\cite{AGL-Spectroscopy}, now including off-diagonal corrections and the presence of several qudits.
It contains a small correction to the eigenstates given by (notice that it has off-diagonal terms $\alpha_1\neq\alpha_2$ which can rotate the original basis):
\begin{equation}
    \delta E_{i,\vec{\alpha}}=\frac{1}{2}\sum_{\beta=1}^{2S+1}\Lambda_{i}^{\alpha_{1},\beta}\Lambda_{i}^{\beta,\alpha_{2}}\left(\frac{1}{E_{i;\alpha_{2},\beta}-\Omega}+\frac{1}{E_{i;\alpha_{1},\beta}-\Omega}\right)
\end{equation}
A state-dependent cavity frequency shift:
\begin{equation}
    \delta \Omega_{i,\vec{\alpha}}=\sum_{\beta=1}^{2S+1}\Lambda_{i}^{\alpha_{1},\beta}\Lambda_{i}^{\beta,\alpha_{2}}\left(\frac{E_{i;\alpha_{1},\beta}}{E_{i;\alpha_{1},\beta}^{2}-\Omega^{2}}+\frac{E_{i;\alpha_{2},\beta}}{E_{i;\alpha_{2},\beta}^{2}-\Omega^{2}}\right)
\end{equation}
which is crucial for read-out protocols using the cavity transmission.
The effective interaction between different qudits is given by:
\begin{equation}
    \tilde{J}_{i,j}^{\vec{\alpha},\vec{\beta}}=\frac{\Omega\Lambda_{i}^{\vec{\beta}}\Lambda_{j}^{\vec{\alpha}}}{E_{j;\vec{\alpha}}^{2}-\Omega^{2}}
    \label{eq:Effective-Interaction-Appendix}
\end{equation}
In addition, the last line in Eq.~\eqref{eq:Effective-Hamiltonian} contains a correction due to two-photon transitions:
\begin{align}
    \tilde{T}_{i,+}^{\vec{\alpha}}=&\frac{1}{2}\sum_{\beta=1}^{2S+1}\Lambda_{i}^{\alpha_{1},\beta}\Lambda_{i}^{\beta,\alpha_{2}}\left(\frac{1}{E_{i;\alpha_{1},\beta}+\Omega}+\frac{1}{E_{i;\alpha_{2},\beta}-\Omega}\right)\\
    \tilde{T}_{i,-}^{\vec{\alpha}}=&\frac{1}{2}\sum_{\beta=1}^{2S+1}\Lambda_{i}^{\alpha_{1},\beta}\Lambda_{i}^{\beta,\alpha_{2}}\left(\frac{1}{E_{i;\alpha_{2},\beta}+\Omega}+\frac{1}{E_{i;\alpha_{1},\beta}-\Omega}\right)
\end{align}
These terms can be neglected for the present case, as the resonator is in its ground state and with a small number of photons.
\section{Full derivation for a pair of molecules with $S=1/2$ coupled to a cavity:}\label{sec:Appendix-B}
Here we fully derive the effective interaction from the S-W transformation and consider the time-evolution operator in the basis of Hubbard operators, to understand more clearly the connection between qubits and qudits.

We start from the standard Hamiltonian for two qubits coupled to a common cavity mode:
\begin{equation}
    H=\sum_{i=L,R}\frac{\Delta_{i}}{2}\sigma_{i}^{z}+\Omega a^\dagger a+\left(a^\dagger + a\right)\sum_{i=L,R}g_x\lambda_i^x\sigma_{i}^{x}
\end{equation}
which can be rewritten in the basis of Hubbard operators as
\begin{equation}
    H_{S}=\sum_{i=L,R}\frac{\Delta_{i}}{2}\left(X_{i}^{+,+}-X_{i}^{-,-}\right)+\Omega a^{\dagger}a+\left(a^{\dagger}+a\right)\sum_{i=L,R}\frac{g_x\lambda_i^x}{2}\left(X_{i}^{+,-}+X_{i}^{-,+}\right)
\end{equation}
By means of the S-W transformation we can write the effective Hamiltonian from Eq.(\ref{eq:Effective-Hamiltonian-Appendix}):
\begin{align}
\tilde{\mathcal{H}}\simeq&\frac{\Omega}{2}\sum_{i=L,R}\frac{\lambda_{i}^{x}g_{x}}{\Delta_{i}^{2}-\Omega^{2}}+\Omega a^{\dagger}a+\sum_{i=L,R}\frac{\Delta_{i}}{2}\left[1+\frac{1}{2}\frac{\left(\lambda_{i}^{x}g_{x}\right)^{2}}{\Delta_{i}^{2}-\Omega^{2}}\right]\left(X_{i}^{+,+}-X_{i}^{-,-}\right)\nonumber\\
&+a^{\dagger}a\sum_{i=L,R}\frac{\Delta_{i}}{2}\frac{\left(\lambda_{i}^{x}g_{x}\right)^{2}}{\Delta_{i}^{2}-\Omega^{2}}\left(X_{i}^{+,+}-X_{i}^{-,-}\right)\nonumber\\
&+\left(a^{\dagger}a^{\dagger}+aa\right)\sum_{i=L,R}\frac{\Delta_{i}}{4}\frac{\left(\lambda_{i}^{x}g_{x}\right)^{2}}{\Delta_{i}^{2}-\Omega^{2}}\left(X_{i}^{+,+}-X_{i}^{-,-}\right)\nonumber\\
&+\frac{1}{4}\sum_{i,j\neq i}\frac{\Omega g_{x}^{2}\lambda_{i}^{x}\lambda_{j}^{x}}{\Delta_{j}^{2}-\Omega^{2}}\left(X_{i}^{+,-}+X_{i}^{-,+}\right)\left(X_{j}^{+,-}+X_{j}^{-,+}\right)
\end{align}
where the last line contains the effective interaction:
\begin{equation}
    \sum_{i,j\neq i}^{N}\sum_{\vec{\alpha},\vec{\beta}=1}^{2S+1}\tilde{J}_{i,j}^{\vec{\alpha},\vec{\beta}}X_{i}^{\vec{\beta}}X_{j}^{\vec{\alpha}} = \frac{\Omega g_{x}^{2}\lambda_{L}^{x}\lambda_{R}^{x}}{4}\left(\frac{1}{\Delta_{R}^{2}-\Omega^{2}}+\frac{1}{\Delta_{L}^{2}-\Omega^{2}}\right)\left(X_{R}^{+,-}+X_{R}^{-,+}\right)\left(X_{L}^{+,-}+X_{L}^{-,+}\right)
\end{equation}
This interaction is of Ising type along the transverse direction, as it will be assumed below for the effective Hamiltonian to describe the dynamics.
To fully identify this Hamiltonian with the one used in the multiple scales analysis, which has spin-spin interactions  only, we need to trace-out the photon sector assuming a density matrix $\rho_p=\sum_{n=0}^{\infty}p_{n}Y^{n,n}$, being $p_n$ the occupation of the state with $n$ photons. 
The final effective Hamiltonain for the two molecules reads (we assume that the cavity is in a well defined number of photons state with zero or one photon):
\begin{align}
    \tilde{\mathcal{H}}\simeq&\frac{\Omega}{2}\sum_{i=L,R}\frac{\lambda_{i}^{x}g_{x}}{\Delta_{i}^{2}-\Omega^{2}}+\Omega p_{1}\nonumber\\
    &+\sum_{i=L,R}\Delta_{i}\left[\frac{1}{2}+\frac{1+2p_{1}}{4}\frac{\left(\lambda_{i}^{x}g_{x}\right)^{2}}{\Delta_{i}^{2}-\Omega^{2}}\right]\left(X_{i}^{+,+}-X_{i}^{-,-}\right)\nonumber\\
    &+\frac{1}{4}\sum_{i,j\neq i}\frac{\Omega g_{x}^{2}\lambda_{i}^{x}\lambda_{j}^{x}}{\Delta_{j}^{2}-\Omega^{2}}\left(X_{i}^{+,-}+X_{i}^{-,+}\right)\left(X_{j}^{+,-}+X_{j}^{-,+}\right)\label{eq:Appendix-Interaction1}
\end{align}
We can see that first line can be ignored, because it represents a global shift in energies. The Zeeman splittings depend on the presence of a photon $p_1$ in the cavity, while the interaction does not. Interestingly, notice how one can tune the sign of the interaction by changing the value of $\Delta_i$ relatively to the cavity frequency $\Omega$.
This expression is the one that can will be used as the effective Hamiltonian in the next section. Furthermore, the dependence on $p_1$ for the qubits energy shift can be ignored for being small in the dispersive regime of operation.
\section{Multiple scales analysis for a pair of two-level systems:}\label{sec:Appendix-C}
Here we review the result from multiple-scales analysis for the case of two qubits interacting via an effective transverse Ising interaction (where $\tilde{V}\ll\tilde{\Delta}_{i}$):
\begin{equation}
    \tilde{\mathcal{H}}=\sum_{i=L,R}\frac{\tilde{\Delta}_{i}}{2}\sigma_{i}^{z}+\tilde{V}\sigma_{L}^{x}\sigma_{R}^{x}\label{eq:Appendix-H1}
\end{equation}
The different terms can be expressed using the microscopic parameters by comparison with Eq.~\eqref{eq:Appendix-Interaction1}.
Concretely, we can identify the effective interaction as:
\begin{equation}
    \tilde{V}=\Omega g_{x}^{2}\lambda_{L}^{x}\lambda_{R}^{x}\left(\frac{1}{\Delta_{R}^{2}-\Omega^{2}}+\frac{1}{\Delta_{L}^{2}-\Omega^{2}}\right)
\end{equation}
and the effective energy for each qubit as:
\begin{equation}
    \tilde{\Delta}_{i} = \Delta_{i}\left[1+\frac{1+2p_{1}}{2}\frac{\left(\lambda_{i}^{x}g_{x}\right)^{2}}{\Delta_{i}^{2}-\Omega^{2}}\right]
\end{equation}

Now we focus on the time evolution operator. We first find the unperturbed solution using multiple scales analysis, and it results in:
\begin{equation}
    U_{0}\left(\vec{\tau}\right)=e^{-i\tau_{0}\sum_{i}\frac{\tilde{\Delta}_{i}}{2}\sigma_{i}^{z}}u_{0}\left(\tau_{1}\right)
\end{equation}
where each qubit freely oscillates with frequency $\Delta_i$. The first order correction from multiple scales analysis requires to solve:
\begin{equation}
    \partial_{\tau_{0}}u_{1}\left(\vec{\tau}\right)=-ie^{i\tilde{\mathcal{H}}_{0}\tau_{0}}\tilde{\mathcal{V}}e^{-i\tilde{\mathcal{H}}_{0}\tau_{0}}u_{0}\left(\tau_{1}\right)-\partial_{\tau_{1}}u_{0}\left(\tau_{1}\right)
\end{equation}
being $U_{1}\left(\vec{\tau}\right)=e^{-i\tau_{0}\tilde{\mathcal{H}}_{0}}u_{1}\left(\vec{\tau}\right)$. The calculation of the first term results yields:
\begin{equation}
    \tilde{V}e^{i\tau\sum_{i}\frac{\tilde{\Delta}_{i}}{2}\sigma_{i}^{z}}\sigma_{L}^{x}\sigma_{R}^{x}e^{-i\tau\sum_{i}\frac{\tilde{\Delta}_{i}}{2}\sigma_{i}^{z}}=\tilde{V}\left[\sigma_{L}^{x}\cos\left(\tilde{\Delta}_{L}\tau\right)+\sigma_{L}^{y}\sin\left(\tilde{\Delta}_{L}\tau\right)\right]\left[\sigma_{R}^{x}\cos\left(\tilde{\Delta}_{R}\tau\right)+\sigma_{R}^{y}\sin\left(\tilde{\Delta}_{R}\tau\right)\right]
\end{equation}
which determines the condition for the presence of secular terms (terms that grow linearly with time and diverge in the asymptotic limit) and must be renormalized if present. 
One can see that for $\tilde{\Delta}_i\geq 0$, the condition $\tilde{\Delta}_L=\tilde{\Delta}_R$ is the one producing these terms. 
Therefore, it is important to distinguish the two cases: i)detuned qubits ($\tilde{\Delta}_L\neq\tilde{\Delta}_R$) and resonant qubits ($\tilde{\Delta}_L=\tilde{\Delta}_R$), which crucially can be controlled externally by means of the local magnetic fields at each qubit. 
This process is what allows to effectively switch on/off the interaction between specific pairs of qubits.

If the qubits are detuned, the solution up to first order in $\tilde{V}$ does not require renormalization and $u_0(\tau_1)$ can be fixed to the identity. Then, the total solution is just:
\begin{equation}
    U(\tau_0)=e^{-i\tau_{0}\sum_{i}\frac{\tilde{\Delta}_{i}}{2}\sigma_{i}^{z}}\left\{ 1-i\tilde{V}\int_{0}^{\tau_{0}}d\tau\left[\sigma_{L}^{x}\cos\left(\tilde{\Delta}_{L}\tau\right)+\sigma_{L}^{y}\sin\left(\tilde{\Delta}_{L}\tau\right)\right]\left[\sigma_{R}^{x}\cos\left(\tilde{\Delta}_{R}\tau\right)+\sigma_{R}^{y}\sin\left(\tilde{\Delta}_{R}\tau\right)\right]\right\} 
\end{equation}
which indicates that, for $\tilde{V}\ll\tilde{\Delta}_i$, the dominant corrections are linear in $\tilde{V}$ and small, and that the qubits freely evolve within their subspace.

In contrast, if we consider the resonant case ($\tilde{\Delta}_L=\tilde{\Delta}_R$), the secular term $\frac{\tilde{V}}{2}\left(\sigma_{L}^{x}\sigma_{R}^{x}+\sigma_{L}^{y}\sigma_{R}^{y}\right)$ must be separated from the first order solution, and cancelled by requiring:
\begin{equation}
    u_{0}\left(\tau_{1}\right)=e^{-i\frac{\tilde{V}}{2}\left(\sigma_{L}^{x}\sigma_{R}^{x}+\sigma_{L}^{y}\sigma_{R}^{y}\right)\tau_{1}}\\=\left(\begin{array}{cccc}
1 & 0 & 0 & 0\\
0 & \cos\left(\tilde{V}\tau_{1}\right) & -i\sin\left(\tilde{V}\tau_{1}\right) & 0\\
0 & -i\sin\left(\tilde{V}\tau_{1}\right) & \cos\left(\tilde{V}\tau_{1}\right) & 0\\
0 & 0 & 0 & 1
\end{array}\right)\label{eq:Appendix-flow1}
\end{equation}
Notice that this makes the lowest order solution $U_0\left(\vec{\tau}\right)$ to drastically change, to encode the secular terms in a non-perturbative way. 
Furthermore, this correction is not small because is not linear in $\tilde{V}$, which indicates that in resonance, the two qubits become highly entangled over a time-scale of $\sim \tilde{V}^{-1}$.
In addition, the unperturbed solution also contains linear corrections in $\tilde{V}$ from the non-secular terms.

To conclude the analysis, let us compare the expression from multiple-scales analysis with the exact expression for the time evolution operator. In resonance ($\tilde{\Delta}_i=\Delta$), we have obtained that the unperturbed solution is given by:
\begin{equation}
U_{0}\left(t\right)	=e^{-it\frac{\Delta}{2}\sum\sigma_{i}^{z}}e^{it\frac{\tilde{V}}{2}\left(\sigma_{L}^{x}\sigma_{R}^{x}+\sigma_{L}^{y}\sigma_{R}^{y}\right)}
U_{0}\left(t\right)	=\left(\begin{array}{cccc}
e^{-i\Delta t} & 0 & 0 & 0\\
0 & \cos\left(\tilde{V}t\right) & -i\sin\left(\tilde{V}t\right) & 0\\
0 & -i\sin\left(\tilde{V}t\right) & \cos\left(\tilde{V}t\right) & 0\\
0 & 0 & 0 & e^{i\Delta t}
\end{array}\right)\label{eq:unperturbed-resonant}
\end{equation}
Then, from exact diagonalization we obtained the following expression for the time-evolution operator:
\begin{equation}
    i \partial_t U\left(t\right) = H U\left(t\right)
\end{equation}
which is a function of the frequencies $\omega_{\pm}=\frac{1}{2}\sqrt{\Delta_{\pm}^{2}+4\tilde{V}^{2}}$ only, with $\Delta_{\pm}=\tilde{\Delta}_{L}-\tilde{\Delta}_{R}$. It is given by the following 4-dimensional matrix in the basis $|\pm,\pm\rangle$:
\begin{equation}
U\left(t\right)=\left(\begin{array}{cccc}
\cos\left(\omega_{+}t\right)-\frac{i\Delta_{+}\sin\left(\omega_{+}t\right)}{2\omega_{+}} & 0 & 0 & -\frac{i\tilde{V}\sin\left(\omega_{+}t\right)}{\omega_{+}}\\
0 & \cos\left(\omega_{-}t\right)-\frac{i\Delta_{-}\sin\left(\omega_{-}t\right)}{2\omega_{-}} & -\frac{i\tilde{V}\sin\left(\omega_{-}t\right)}{\omega_{-}} & 0\\
0 & -\frac{i\tilde{V}\sin\left(\omega_{-}t\right)}{\omega_{-}} & \cos\left(\omega_{-}t\right)+\frac{i\Delta_{-}\sin\left(\omega_{-}t\right)}{2\omega_{-}} & 0\\
-\frac{i\tilde{V}\sin\left(\omega_{+}t\right)}{\omega_{+}} & 0 & 0 & \cos\left(\omega_{+}t\right)+\frac{i\Delta_{+}\sin\left(\omega_{+}t\right)}{2\omega_{+}}
\end{array}\right)
\label{eq:Matrix1}
\end{equation}
As it can be seen, for $\tilde{\Delta}_L\neq\tilde{\Delta}_R$, $\tilde{\Delta}_{i}\geq 0$ and $\tilde{\Delta}_{i}\gg \tilde{V}$, the off-diagonal terms in Eq.\eqref{eq:Matrix1}, proportional to $\tilde{V}$ are very small, which indicates that each qubit freely evolves within its own subspace. In contrast, when the qubits are in the resonant condition $\tilde{\Delta}_i=\Delta$ the time-evolution operator simplifies to ($\omega_0=\sqrt{\Delta^{2}+\tilde{V}^{2}}$):
\begin{equation}
U\left(t\right)\underset{\tilde{\Delta}_{j}=\Delta}{\longrightarrow}\left(\begin{array}{cccc}
\cos\left(\omega_0 t\right)-i\frac{\Delta}{\omega_0}\sin\left(\omega_0 t\right) & 0 & 0 & -i\frac{\tilde{V}}{\omega_0}\sin\left(\omega_0 t\right)\\
0 & \cos\left(\tilde{V}t\right) & -i\sin\left(\tilde{V}t\right) & 0\\
0 & -i\sin\left(\tilde{V}t\right) & \cos\left(\tilde{V}t\right) & 0\\
-i\frac{\tilde{V}}{\omega_0}\sin\left(\omega_0 t\right) & 0 & 0 & \cos\left(\omega t\right)+i\frac{\Delta}{\omega_0}\sin\left(\omega_0 t\right)
\end{array}\right)
\end{equation}
which agrees with our result from multiple-scales analysis, if we assume that $\omega_0\sim\Delta$, which is the case for our setup with $\tilde{V}\ll\Delta$. The off-diagonal terms missing in Eq.\eqref{eq:unperturbed-resonant}, linear in $\tilde{V}$, are obtained when the first order corrections in $\tilde{V}$ from non-secular terms are added.

It is important to notice that the time-evolution operator can be used to produce an iSWAP gate if we allow the qubits interact for a time $t=\pi/2\tilde{V}$. 
Engineering other types of gates would require to start from a different effective interaction or to apply additional time-dependent protocols $\tilde{V}(t)$ which would allow to change the secular terms.


\end{widetext}
\bibliographystyle{apsrev4-2}
\bibliography{bibliography}

\begin{thebibliography}{55}%
\makeatletter
\providecommand \@ifxundefined [1]{%
 \@ifx{#1\undefined}
}%
\providecommand \@ifnum [1]{%
 \ifnum #1\expandafter \@firstoftwo
 \else \expandafter \@secondoftwo
 \fi
}%
\providecommand \@ifx [1]{%
 \ifx #1\expandafter \@firstoftwo
 \else \expandafter \@secondoftwo
 \fi
}%
\providecommand \natexlab [1]{#1}%
\providecommand \enquote  [1]{``#1''}%
\providecommand \bibnamefont  [1]{#1}%
\providecommand \bibfnamefont [1]{#1}%
\providecommand \citenamefont [1]{#1}%
\providecommand \href@noop [0]{\@secondoftwo}%
\providecommand \href [0]{\begingroup \@sanitize@url \@href}%
\providecommand \@href[1]{\@@startlink{#1}\@@href}%
\providecommand \@@href[1]{\endgroup#1\@@endlink}%
\providecommand \@sanitize@url [0]{\catcode `\\12\catcode `\$12\catcode
  `\&12\catcode `\#12\catcode `\^12\catcode `\_12\catcode `\%12\relax}%
\providecommand \@@startlink[1]{}%
\providecommand \@@endlink[0]{}%
\providecommand \url  [0]{\begingroup\@sanitize@url \@url }%
\providecommand \@url [1]{\endgroup\@href {#1}{\urlprefix }}%
\providecommand \urlprefix  [0]{URL }%
\providecommand \Eprint [0]{\href }%
\providecommand \doibase [0]{https://doi.org/}%
\providecommand \selectlanguage [0]{\@gobble}%
\providecommand \bibinfo  [0]{\@secondoftwo}%
\providecommand \bibfield  [0]{\@secondoftwo}%
\providecommand \translation [1]{[#1]}%
\providecommand \BibitemOpen [0]{}%
\providecommand \bibitemStop [0]{}%
\providecommand \bibitemNoStop [0]{.\EOS\space}%
\providecommand \EOS [0]{\spacefactor3000\relax}%
\providecommand \BibitemShut  [1]{\csname bibitem#1\endcsname}%
\let\auto@bib@innerbib\@empty
\bibitem [{\citenamefont {Kimble}(2008)}]{QuantumInternet}%
  \BibitemOpen
  \bibfield  {author} {\bibinfo {author} {\bibfnamefont {H.~J.}\ \bibnamefont
  {Kimble}},\ }\href {https://doi.org/10.1038/nature07127} {\bibfield
  {journal} {\bibinfo  {journal} {Nature}\ }\textbf {\bibinfo {volume} {453}},\
  \bibinfo {pages} {1023} (\bibinfo {year} {2008})}\BibitemShut {NoStop}%
\bibitem [{\citenamefont {Zinner}\ \emph {et~al.}(2021)\citenamefont {Zinner},
  \citenamefont {Dahlhausen}, \citenamefont {Boehme}, \citenamefont {Ehlers},
  \citenamefont {Bieske},\ and\ \citenamefont {Fehring}}]{DrugDiscovery}%
  \BibitemOpen
  \bibfield  {author} {\bibinfo {author} {\bibfnamefont {M.}~\bibnamefont
  {Zinner}}, \bibinfo {author} {\bibfnamefont {F.}~\bibnamefont {Dahlhausen}},
  \bibinfo {author} {\bibfnamefont {P.}~\bibnamefont {Boehme}}, \bibinfo
  {author} {\bibfnamefont {J.}~\bibnamefont {Ehlers}}, \bibinfo {author}
  {\bibfnamefont {L.}~\bibnamefont {Bieske}},\ and\ \bibinfo {author}
  {\bibfnamefont {L.}~\bibnamefont {Fehring}},\ }\href
  {https://doi.org/https://doi.org/10.1016/j.drudis.2021.06.003} {\bibfield
  {journal} {\bibinfo  {journal} {Drug Discovery Today}\ }\textbf {\bibinfo
  {volume} {26}},\ \bibinfo {pages} {1680} (\bibinfo {year}
  {2021})}\BibitemShut {NoStop}%
\bibitem [{\citenamefont {Louie}\ \emph {et~al.}(2021)\citenamefont {Louie},
  \citenamefont {Chan}, \citenamefont {da~Jornada}, \citenamefont {Li},\ and\
  \citenamefont {Qiu}}]{NewMaterials}%
  \BibitemOpen
  \bibfield  {author} {\bibinfo {author} {\bibfnamefont {S.~G.}\ \bibnamefont
  {Louie}}, \bibinfo {author} {\bibfnamefont {Y.-H.}\ \bibnamefont {Chan}},
  \bibinfo {author} {\bibfnamefont {F.~H.}\ \bibnamefont {da~Jornada}},
  \bibinfo {author} {\bibfnamefont {Z.}~\bibnamefont {Li}},\ and\ \bibinfo
  {author} {\bibfnamefont {D.~Y.}\ \bibnamefont {Qiu}},\ }\href
  {https://doi.org/10.1038/s41563-021-01015-1} {\bibfield  {journal} {\bibinfo
  {journal} {Nature Materials}\ }\textbf {\bibinfo {volume} {20}},\ \bibinfo
  {pages} {728} (\bibinfo {year} {2021})}\BibitemShut {NoStop}%
\bibitem [{\citenamefont {Preskill}(2018)}]{Preskill2018quantumcomputingin}%
  \BibitemOpen
  \bibfield  {author} {\bibinfo {author} {\bibfnamefont {J.}~\bibnamefont
  {Preskill}},\ }\href {https://doi.org/10.22331/q-2018-08-06-79} {\bibfield
  {journal} {\bibinfo  {journal} {{Quantum}}\ }\textbf {\bibinfo {volume}
  {2}},\ \bibinfo {pages} {79} (\bibinfo {year} {2018})}\BibitemShut {NoStop}%
\bibitem [{\citenamefont {Arute}\ \emph {et~al.}(2019)\citenamefont {Arute},
  \citenamefont {Arya}, \citenamefont {Babbush}, \citenamefont {Bacon},
  \citenamefont {Bardin}, \citenamefont {Barends}, \citenamefont {Biswas},
  \citenamefont {Boixo}, \citenamefont {Brandao}, \citenamefont {Buell},
  \citenamefont {Burkett}, \citenamefont {Chen}, \citenamefont {Chen},
  \citenamefont {Chiaro}, \citenamefont {Collins}, \citenamefont {Courtney},
  \citenamefont {Dunsworth}, \citenamefont {Farhi}, \citenamefont {Foxen},
  \citenamefont {Fowler}, \citenamefont {Gidney}, \citenamefont {Giustina},
  \citenamefont {Graff}, \citenamefont {Guerin}, \citenamefont {Habegger},
  \citenamefont {Harrigan}, \citenamefont {Hartmann}, \citenamefont {Ho},
  \citenamefont {Hoffmann}, \citenamefont {Huang}, \citenamefont {Humble},
  \citenamefont {Isakov}, \citenamefont {Jeffrey}, \citenamefont {Jiang},
  \citenamefont {Kafri}, \citenamefont {Kechedzhi}, \citenamefont {Kelly},
  \citenamefont {Klimov}, \citenamefont {Knysh}, \citenamefont {Korotkov},
  \citenamefont {Kostritsa}, \citenamefont {Landhuis}, \citenamefont
  {Lindmark}, \citenamefont {Lucero}, \citenamefont {Lyakh}, \citenamefont
  {Mandr{\`a}}, \citenamefont {McClean}, \citenamefont {McEwen}, \citenamefont
  {Megrant}, \citenamefont {Mi}, \citenamefont {Michielsen}, \citenamefont
  {Mohseni}, \citenamefont {Mutus}, \citenamefont {Naaman}, \citenamefont
  {Neeley}, \citenamefont {Neill}, \citenamefont {Niu}, \citenamefont {Ostby},
  \citenamefont {Petukhov}, \citenamefont {Platt}, \citenamefont {Quintana},
  \citenamefont {Rieffel}, \citenamefont {Roushan}, \citenamefont {Rubin},
  \citenamefont {Sank}, \citenamefont {Satzinger}, \citenamefont {Smelyanskiy},
  \citenamefont {Sung}, \citenamefont {Trevithick}, \citenamefont
  {Vainsencher}, \citenamefont {Villalonga}, \citenamefont {White},
  \citenamefont {Yao}, \citenamefont {Yeh}, \citenamefont {Zalcman},
  \citenamefont {Neven},\ and\ \citenamefont {Martinis}}]{QSupremacy}%
  \BibitemOpen
  \bibfield  {author} {\bibinfo {author} {\bibfnamefont {F.}~\bibnamefont
  {Arute}}, \bibinfo {author} {\bibfnamefont {K.}~\bibnamefont {Arya}},
  \bibinfo {author} {\bibfnamefont {R.}~\bibnamefont {Babbush}}, \bibinfo
  {author} {\bibfnamefont {D.}~\bibnamefont {Bacon}}, \bibinfo {author}
  {\bibfnamefont {J.~C.}\ \bibnamefont {Bardin}}, \bibinfo {author}
  {\bibfnamefont {R.}~\bibnamefont {Barends}}, \bibinfo {author} {\bibfnamefont
  {R.}~\bibnamefont {Biswas}}, \bibinfo {author} {\bibfnamefont
  {S.}~\bibnamefont {Boixo}}, \bibinfo {author} {\bibfnamefont {F.~G. S.~L.}\
  \bibnamefont {Brandao}}, \bibinfo {author} {\bibfnamefont {D.~A.}\
  \bibnamefont {Buell}}, \bibinfo {author} {\bibfnamefont {B.}~\bibnamefont
  {Burkett}}, \bibinfo {author} {\bibfnamefont {Y.}~\bibnamefont {Chen}},
  \bibinfo {author} {\bibfnamefont {Z.}~\bibnamefont {Chen}}, \bibinfo {author}
  {\bibfnamefont {B.}~\bibnamefont {Chiaro}}, \bibinfo {author} {\bibfnamefont
  {R.}~\bibnamefont {Collins}}, \bibinfo {author} {\bibfnamefont
  {W.}~\bibnamefont {Courtney}}, \bibinfo {author} {\bibfnamefont
  {A.}~\bibnamefont {Dunsworth}}, \bibinfo {author} {\bibfnamefont
  {E.}~\bibnamefont {Farhi}}, \bibinfo {author} {\bibfnamefont
  {B.}~\bibnamefont {Foxen}}, \bibinfo {author} {\bibfnamefont
  {A.}~\bibnamefont {Fowler}}, \bibinfo {author} {\bibfnamefont
  {C.}~\bibnamefont {Gidney}}, \bibinfo {author} {\bibfnamefont
  {M.}~\bibnamefont {Giustina}}, \bibinfo {author} {\bibfnamefont
  {R.}~\bibnamefont {Graff}}, \bibinfo {author} {\bibfnamefont
  {K.}~\bibnamefont {Guerin}}, \bibinfo {author} {\bibfnamefont
  {S.}~\bibnamefont {Habegger}}, \bibinfo {author} {\bibfnamefont {M.~P.}\
  \bibnamefont {Harrigan}}, \bibinfo {author} {\bibfnamefont {M.~J.}\
  \bibnamefont {Hartmann}}, \bibinfo {author} {\bibfnamefont {A.}~\bibnamefont
  {Ho}}, \bibinfo {author} {\bibfnamefont {M.}~\bibnamefont {Hoffmann}},
  \bibinfo {author} {\bibfnamefont {T.}~\bibnamefont {Huang}}, \bibinfo
  {author} {\bibfnamefont {T.~S.}\ \bibnamefont {Humble}}, \bibinfo {author}
  {\bibfnamefont {S.~V.}\ \bibnamefont {Isakov}}, \bibinfo {author}
  {\bibfnamefont {E.}~\bibnamefont {Jeffrey}}, \bibinfo {author} {\bibfnamefont
  {Z.}~\bibnamefont {Jiang}}, \bibinfo {author} {\bibfnamefont
  {D.}~\bibnamefont {Kafri}}, \bibinfo {author} {\bibfnamefont
  {K.}~\bibnamefont {Kechedzhi}}, \bibinfo {author} {\bibfnamefont
  {J.}~\bibnamefont {Kelly}}, \bibinfo {author} {\bibfnamefont {P.~V.}\
  \bibnamefont {Klimov}}, \bibinfo {author} {\bibfnamefont {S.}~\bibnamefont
  {Knysh}}, \bibinfo {author} {\bibfnamefont {A.}~\bibnamefont {Korotkov}},
  \bibinfo {author} {\bibfnamefont {F.}~\bibnamefont {Kostritsa}}, \bibinfo
  {author} {\bibfnamefont {D.}~\bibnamefont {Landhuis}}, \bibinfo {author}
  {\bibfnamefont {M.}~\bibnamefont {Lindmark}}, \bibinfo {author}
  {\bibfnamefont {E.}~\bibnamefont {Lucero}}, \bibinfo {author} {\bibfnamefont
  {D.}~\bibnamefont {Lyakh}}, \bibinfo {author} {\bibfnamefont
  {S.}~\bibnamefont {Mandr{\`a}}}, \bibinfo {author} {\bibfnamefont {J.~R.}\
  \bibnamefont {McClean}}, \bibinfo {author} {\bibfnamefont {M.}~\bibnamefont
  {McEwen}}, \bibinfo {author} {\bibfnamefont {A.}~\bibnamefont {Megrant}},
  \bibinfo {author} {\bibfnamefont {X.}~\bibnamefont {Mi}}, \bibinfo {author}
  {\bibfnamefont {K.}~\bibnamefont {Michielsen}}, \bibinfo {author}
  {\bibfnamefont {M.}~\bibnamefont {Mohseni}}, \bibinfo {author} {\bibfnamefont
  {J.}~\bibnamefont {Mutus}}, \bibinfo {author} {\bibfnamefont
  {O.}~\bibnamefont {Naaman}}, \bibinfo {author} {\bibfnamefont
  {M.}~\bibnamefont {Neeley}}, \bibinfo {author} {\bibfnamefont
  {C.}~\bibnamefont {Neill}}, \bibinfo {author} {\bibfnamefont {M.~Y.}\
  \bibnamefont {Niu}}, \bibinfo {author} {\bibfnamefont {E.}~\bibnamefont
  {Ostby}}, \bibinfo {author} {\bibfnamefont {A.}~\bibnamefont {Petukhov}},
  \bibinfo {author} {\bibfnamefont {J.~C.}\ \bibnamefont {Platt}}, \bibinfo
  {author} {\bibfnamefont {C.}~\bibnamefont {Quintana}}, \bibinfo {author}
  {\bibfnamefont {E.~G.}\ \bibnamefont {Rieffel}}, \bibinfo {author}
  {\bibfnamefont {P.}~\bibnamefont {Roushan}}, \bibinfo {author} {\bibfnamefont
  {N.~C.}\ \bibnamefont {Rubin}}, \bibinfo {author} {\bibfnamefont
  {D.}~\bibnamefont {Sank}}, \bibinfo {author} {\bibfnamefont {K.~J.}\
  \bibnamefont {Satzinger}}, \bibinfo {author} {\bibfnamefont {V.}~\bibnamefont
  {Smelyanskiy}}, \bibinfo {author} {\bibfnamefont {K.~J.}\ \bibnamefont
  {Sung}}, \bibinfo {author} {\bibfnamefont {M.~D.}\ \bibnamefont
  {Trevithick}}, \bibinfo {author} {\bibfnamefont {A.}~\bibnamefont
  {Vainsencher}}, \bibinfo {author} {\bibfnamefont {B.}~\bibnamefont
  {Villalonga}}, \bibinfo {author} {\bibfnamefont {T.}~\bibnamefont {White}},
  \bibinfo {author} {\bibfnamefont {Z.~J.}\ \bibnamefont {Yao}}, \bibinfo
  {author} {\bibfnamefont {P.}~\bibnamefont {Yeh}}, \bibinfo {author}
  {\bibfnamefont {A.}~\bibnamefont {Zalcman}}, \bibinfo {author} {\bibfnamefont
  {H.}~\bibnamefont {Neven}},\ and\ \bibinfo {author} {\bibfnamefont {J.~M.}\
  \bibnamefont {Martinis}},\ }\href {https://doi.org/10.1038/s41586-019-1666-5}
  {\bibfield  {journal} {\bibinfo  {journal} {Nature}\ }\textbf {\bibinfo
  {volume} {574}},\ \bibinfo {pages} {505} (\bibinfo {year}
  {2019})}\BibitemShut {NoStop}%
\bibitem [{\citenamefont {Kielpinski}\ \emph {et~al.}(2002)\citenamefont
  {Kielpinski}, \citenamefont {Monroe},\ and\ \citenamefont
  {Wineland}}]{QC-Ions0}%
  \BibitemOpen
  \bibfield  {author} {\bibinfo {author} {\bibfnamefont {D.}~\bibnamefont
  {Kielpinski}}, \bibinfo {author} {\bibfnamefont {C.}~\bibnamefont {Monroe}},\
  and\ \bibinfo {author} {\bibfnamefont {D.~J.}\ \bibnamefont {Wineland}},\
  }\href {https://doi.org/10.1038/nature00784} {\bibfield  {journal} {\bibinfo
  {journal} {Nature}\ }\textbf {\bibinfo {volume} {417}},\ \bibinfo {pages}
  {709} (\bibinfo {year} {2002})}\BibitemShut {NoStop}%
\bibitem [{\citenamefont {Garc\'{\i}a-Ripoll}\ \emph
  {et~al.}(2003)\citenamefont {Garc\'{\i}a-Ripoll}, \citenamefont {Zoller},\
  and\ \citenamefont {Cirac}}]{QC-Ions1}%
  \BibitemOpen
  \bibfield  {author} {\bibinfo {author} {\bibfnamefont {J.~J.}\ \bibnamefont
  {Garc\'{\i}a-Ripoll}}, \bibinfo {author} {\bibfnamefont {P.}~\bibnamefont
  {Zoller}},\ and\ \bibinfo {author} {\bibfnamefont {J.~I.}\ \bibnamefont
  {Cirac}},\ }\href {https://doi.org/10.1103/PhysRevLett.91.157901} {\bibfield
  {journal} {\bibinfo  {journal} {Phys. Rev. Lett.}\ }\textbf {\bibinfo
  {volume} {91}},\ \bibinfo {pages} {157901} (\bibinfo {year}
  {2003})}\BibitemShut {NoStop}%
\bibitem [{\citenamefont {Pogorelov}\ \emph {et~al.}(2021)\citenamefont
  {Pogorelov}, \citenamefont {Feldker}, \citenamefont {Marciniak},
  \citenamefont {Postler}, \citenamefont {Jacob}, \citenamefont
  {Krieglsteiner}, \citenamefont {Podlesnic}, \citenamefont {Meth},
  \citenamefont {Negnevitsky}, \citenamefont {Stadler}, \citenamefont
  {H\"ofer}, \citenamefont {W\"achter}, \citenamefont {Lakhmanskiy},
  \citenamefont {Blatt}, \citenamefont {Schindler},\ and\ \citenamefont
  {Monz}}]{QC-Ions2}%
  \BibitemOpen
  \bibfield  {author} {\bibinfo {author} {\bibfnamefont {I.}~\bibnamefont
  {Pogorelov}}, \bibinfo {author} {\bibfnamefont {T.}~\bibnamefont {Feldker}},
  \bibinfo {author} {\bibfnamefont {C.~D.}\ \bibnamefont {Marciniak}}, \bibinfo
  {author} {\bibfnamefont {L.}~\bibnamefont {Postler}}, \bibinfo {author}
  {\bibfnamefont {G.}~\bibnamefont {Jacob}}, \bibinfo {author} {\bibfnamefont
  {O.}~\bibnamefont {Krieglsteiner}}, \bibinfo {author} {\bibfnamefont
  {V.}~\bibnamefont {Podlesnic}}, \bibinfo {author} {\bibfnamefont
  {M.}~\bibnamefont {Meth}}, \bibinfo {author} {\bibfnamefont {V.}~\bibnamefont
  {Negnevitsky}}, \bibinfo {author} {\bibfnamefont {M.}~\bibnamefont
  {Stadler}}, \bibinfo {author} {\bibfnamefont {B.}~\bibnamefont {H\"ofer}},
  \bibinfo {author} {\bibfnamefont {C.}~\bibnamefont {W\"achter}}, \bibinfo
  {author} {\bibfnamefont {K.}~\bibnamefont {Lakhmanskiy}}, \bibinfo {author}
  {\bibfnamefont {R.}~\bibnamefont {Blatt}}, \bibinfo {author} {\bibfnamefont
  {P.}~\bibnamefont {Schindler}},\ and\ \bibinfo {author} {\bibfnamefont
  {T.}~\bibnamefont {Monz}},\ }\href
  {https://doi.org/10.1103/PRXQuantum.2.020343} {\bibfield  {journal} {\bibinfo
   {journal} {PRX Quantum}\ }\textbf {\bibinfo {volume} {2}},\ \bibinfo {pages}
  {020343} (\bibinfo {year} {2021})}\BibitemShut {NoStop}%
\bibitem [{\citenamefont {Clarke}\ and\ \citenamefont
  {Wilhelm}(2008)}]{QC-Supercond0}%
  \BibitemOpen
  \bibfield  {author} {\bibinfo {author} {\bibfnamefont {J.}~\bibnamefont
  {Clarke}}\ and\ \bibinfo {author} {\bibfnamefont {F.~K.}\ \bibnamefont
  {Wilhelm}},\ }\href@noop {} {\bibfield  {journal} {\bibinfo  {journal}
  {Nature}\ }\textbf {\bibinfo {volume} {453}},\ \bibinfo {pages} {1031}
  (\bibinfo {year} {2008})}\BibitemShut {NoStop}%
\bibitem [{\citenamefont {Imamog\ifmmode\bar\else\textasciimacron\fi{}lu}\
  \emph {et~al.}(1999)\citenamefont
  {Imamog\ifmmode\bar\else\textasciimacron\fi{}lu}, \citenamefont {Awschalom},
  \citenamefont {Burkard}, \citenamefont {DiVincenzo}, \citenamefont {Loss},
  \citenamefont {Sherwin},\ and\ \citenamefont {Small}}]{QC-QDots0}%
  \BibitemOpen
  \bibfield  {author} {\bibinfo {author} {\bibfnamefont {A.}~\bibnamefont
  {Imamog\ifmmode\bar\else\textasciimacron\fi{}lu}}, \bibinfo {author}
  {\bibfnamefont {D.~D.}\ \bibnamefont {Awschalom}}, \bibinfo {author}
  {\bibfnamefont {G.}~\bibnamefont {Burkard}}, \bibinfo {author} {\bibfnamefont
  {D.~P.}\ \bibnamefont {DiVincenzo}}, \bibinfo {author} {\bibfnamefont
  {D.}~\bibnamefont {Loss}}, \bibinfo {author} {\bibfnamefont {M.}~\bibnamefont
  {Sherwin}},\ and\ \bibinfo {author} {\bibfnamefont {A.}~\bibnamefont
  {Small}},\ }\href {https://doi.org/10.1103/PhysRevLett.83.4204} {\bibfield
  {journal} {\bibinfo  {journal} {Phys. Rev. Lett.}\ }\textbf {\bibinfo
  {volume} {83}},\ \bibinfo {pages} {4204} (\bibinfo {year}
  {1999})}\BibitemShut {NoStop}%
\bibitem [{\citenamefont {Zhong}\ \emph {et~al.}(2020)\citenamefont {Zhong},
  \citenamefont {Wang}, \citenamefont {Deng}, \citenamefont {Chen},
  \citenamefont {Peng}, \citenamefont {Luo}, \citenamefont {Qin}, \citenamefont
  {Wu}, \citenamefont {Ding}, \citenamefont {Hu}, \citenamefont {Hu},
  \citenamefont {Yang}, \citenamefont {Zhang}, \citenamefont {Li},
  \citenamefont {Li}, \citenamefont {Jiang}, \citenamefont {Gan}, \citenamefont
  {Yang}, \citenamefont {You}, \citenamefont {Wang}, \citenamefont {Li},
  \citenamefont {Liu}, \citenamefont {Lu},\ and\ \citenamefont
  {Pan}}]{BosonSampling}%
  \BibitemOpen
  \bibfield  {author} {\bibinfo {author} {\bibfnamefont {H.-S.}\ \bibnamefont
  {Zhong}}, \bibinfo {author} {\bibfnamefont {H.}~\bibnamefont {Wang}},
  \bibinfo {author} {\bibfnamefont {Y.-H.}\ \bibnamefont {Deng}}, \bibinfo
  {author} {\bibfnamefont {M.-C.}\ \bibnamefont {Chen}}, \bibinfo {author}
  {\bibfnamefont {L.-C.}\ \bibnamefont {Peng}}, \bibinfo {author}
  {\bibfnamefont {Y.-H.}\ \bibnamefont {Luo}}, \bibinfo {author} {\bibfnamefont
  {J.}~\bibnamefont {Qin}}, \bibinfo {author} {\bibfnamefont {D.}~\bibnamefont
  {Wu}}, \bibinfo {author} {\bibfnamefont {X.}~\bibnamefont {Ding}}, \bibinfo
  {author} {\bibfnamefont {Y.}~\bibnamefont {Hu}}, \bibinfo {author}
  {\bibfnamefont {P.}~\bibnamefont {Hu}}, \bibinfo {author} {\bibfnamefont
  {X.-Y.}\ \bibnamefont {Yang}}, \bibinfo {author} {\bibfnamefont {W.-J.}\
  \bibnamefont {Zhang}}, \bibinfo {author} {\bibfnamefont {H.}~\bibnamefont
  {Li}}, \bibinfo {author} {\bibfnamefont {Y.}~\bibnamefont {Li}}, \bibinfo
  {author} {\bibfnamefont {X.}~\bibnamefont {Jiang}}, \bibinfo {author}
  {\bibfnamefont {L.}~\bibnamefont {Gan}}, \bibinfo {author} {\bibfnamefont
  {G.}~\bibnamefont {Yang}}, \bibinfo {author} {\bibfnamefont {L.}~\bibnamefont
  {You}}, \bibinfo {author} {\bibfnamefont {Z.}~\bibnamefont {Wang}}, \bibinfo
  {author} {\bibfnamefont {L.}~\bibnamefont {Li}}, \bibinfo {author}
  {\bibfnamefont {N.-L.}\ \bibnamefont {Liu}}, \bibinfo {author} {\bibfnamefont
  {C.-Y.}\ \bibnamefont {Lu}},\ and\ \bibinfo {author} {\bibfnamefont {J.-W.}\
  \bibnamefont {Pan}},\ }\href {https://doi.org/10.1126/science.abe8770}
  {\bibfield  {journal} {\bibinfo  {journal} {Science}\ }\textbf {\bibinfo
  {volume} {370}},\ \bibinfo {pages} {1460} (\bibinfo {year} {2020})},\ \Eprint
  {https://arxiv.org/abs/https://www.science.org/doi/pdf/10.1126/science.abe8770}
  {https://www.science.org/doi/pdf/10.1126/science.abe8770} \BibitemShut
  {NoStop}%
\bibitem [{\citenamefont {Leuenberger}\ and\ \citenamefont
  {Loss}(2001)}]{QC-Magnets0}%
  \BibitemOpen
  \bibfield  {author} {\bibinfo {author} {\bibfnamefont {M.~N.}\ \bibnamefont
  {Leuenberger}}\ and\ \bibinfo {author} {\bibfnamefont {D.}~\bibnamefont
  {Loss}},\ }\href {https://doi.org/10.1038/35071024} {\bibfield  {journal}
  {\bibinfo  {journal} {Nature}\ }\textbf {\bibinfo {volume} {410}},\ \bibinfo
  {pages} {789} (\bibinfo {year} {2001})}\BibitemShut {NoStop}%
\bibitem [{\citenamefont {Gaita-Ari\~{n}o}\ \emph {et~al.}(2019)\citenamefont
  {Gaita-Ari\~{n}o}, \citenamefont {Luis}, \citenamefont {Hill},\ and\
  \citenamefont {Coronado}}]{FLuis-MolecularSpins&QC}%
  \BibitemOpen
  \bibfield  {author} {\bibinfo {author} {\bibfnamefont {A.}~\bibnamefont
  {Gaita-Ari\~{n}o}}, \bibinfo {author} {\bibfnamefont {F.}~\bibnamefont
  {Luis}}, \bibinfo {author} {\bibfnamefont {S.}~\bibnamefont {Hill}},\ and\
  \bibinfo {author} {\bibfnamefont {E.}~\bibnamefont {Coronado}},\ }\href
  {https://doi.org/10.1038/s41557-019-0232-y} {\bibfield  {journal} {\bibinfo
  {journal} {Nature Chemistry}\ }\textbf {\bibinfo {volume} {11}},\ \bibinfo
  {pages} {301} (\bibinfo {year} {2019})}\BibitemShut {NoStop}%
\bibitem [{\citenamefont {Jenkins}\ \emph {et~al.}(2016)\citenamefont
  {Jenkins}, \citenamefont {Zueco}, \citenamefont {Roubeau}, \citenamefont
  {Arom{\'{i}}}, \citenamefont {Majer},\ and\ \citenamefont
  {Luis}}]{FLuis-ScalableArchitecture}%
  \BibitemOpen
  \bibfield  {author} {\bibinfo {author} {\bibfnamefont {M.~D.}\ \bibnamefont
  {Jenkins}}, \bibinfo {author} {\bibfnamefont {D.}~\bibnamefont {Zueco}},
  \bibinfo {author} {\bibfnamefont {O.}~\bibnamefont {Roubeau}}, \bibinfo
  {author} {\bibfnamefont {G.}~\bibnamefont {Arom{\'{i}}}}, \bibinfo {author}
  {\bibfnamefont {J.}~\bibnamefont {Majer}},\ and\ \bibinfo {author}
  {\bibfnamefont {F.}~\bibnamefont {Luis}},\ }\href
  {https://doi.org/10.1039/C6DT02664H} {\bibfield  {journal} {\bibinfo
  {journal} {Dalton Trans.}\ }\textbf {\bibinfo {volume} {45}},\ \bibinfo
  {pages} {16682} (\bibinfo {year} {2016})}\BibitemShut {NoStop}%
\bibitem [{\citenamefont {Coronado}(2020)}]{Coronado2020}%
  \BibitemOpen
  \bibfield  {author} {\bibinfo {author} {\bibfnamefont {E.}~\bibnamefont
  {Coronado}},\ }\href {https://doi.org/10.1038/s41578-019-0146-8} {\bibfield
  {journal} {\bibinfo  {journal} {Nature Reviews Materials}\ }\textbf {\bibinfo
  {volume} {5}},\ \bibinfo {pages} {87} (\bibinfo {year} {2020})}\BibitemShut
  {NoStop}%
\bibitem [{\citenamefont {Gimeno}\ \emph {et~al.}(2021)\citenamefont {Gimeno},
  \citenamefont {Urtizberea}, \citenamefont {Rom\'an-Roche}, \citenamefont
  {Zueco}, \citenamefont {Camón}, \citenamefont {Alonso}, \citenamefont
  {Roubeau},\ and\ \citenamefont {Luis}}]{Vanadyl1}%
  \BibitemOpen
  \bibfield  {author} {\bibinfo {author} {\bibfnamefont {I.}~\bibnamefont
  {Gimeno}}, \bibinfo {author} {\bibfnamefont {A.}~\bibnamefont {Urtizberea}},
  \bibinfo {author} {\bibfnamefont {J.}~\bibnamefont {Rom\'an-Roche}}, \bibinfo
  {author} {\bibfnamefont {D.}~\bibnamefont {Zueco}}, \bibinfo {author}
  {\bibfnamefont {A.}~\bibnamefont {Camón}}, \bibinfo {author} {\bibfnamefont
  {P.~J.}\ \bibnamefont {Alonso}}, \bibinfo {author} {\bibfnamefont
  {O.}~\bibnamefont {Roubeau}},\ and\ \bibinfo {author} {\bibfnamefont
  {F.}~\bibnamefont {Luis}},\ }\href {https://doi.org/10.1039/D1SC00564B}
  {\bibfield  {journal} {\bibinfo  {journal} {Chem. Sci.}\ }\textbf {\bibinfo
  {volume} {12}},\ \bibinfo {pages} {5621} (\bibinfo {year}
  {2021})}\BibitemShut {NoStop}%
\bibitem [{\citenamefont {Chicco}\ \emph {et~al.}(2021)\citenamefont {Chicco},
  \citenamefont {Chiesa}, \citenamefont {Allodi}, \citenamefont {Garlatti},
  \citenamefont {Atzori}, \citenamefont {Sorace}, \citenamefont {De~Renzi},
  \citenamefont {Sessoli},\ and\ \citenamefont {Carretta}}]{Vanadyl2}%
  \BibitemOpen
  \bibfield  {author} {\bibinfo {author} {\bibfnamefont {S.}~\bibnamefont
  {Chicco}}, \bibinfo {author} {\bibfnamefont {A.}~\bibnamefont {Chiesa}},
  \bibinfo {author} {\bibfnamefont {G.}~\bibnamefont {Allodi}}, \bibinfo
  {author} {\bibfnamefont {E.}~\bibnamefont {Garlatti}}, \bibinfo {author}
  {\bibfnamefont {M.}~\bibnamefont {Atzori}}, \bibinfo {author} {\bibfnamefont
  {L.}~\bibnamefont {Sorace}}, \bibinfo {author} {\bibfnamefont
  {R.}~\bibnamefont {De~Renzi}}, \bibinfo {author} {\bibfnamefont
  {R.}~\bibnamefont {Sessoli}},\ and\ \bibinfo {author} {\bibfnamefont
  {S.}~\bibnamefont {Carretta}},\ }\href {https://doi.org/10.1039/D1SC01358K}
  {\bibfield  {journal} {\bibinfo  {journal} {Chem. Sci.}\ }\textbf {\bibinfo
  {volume} {12}},\ \bibinfo {pages} {12046} (\bibinfo {year}
  {2021})}\BibitemShut {NoStop}%
\bibitem [{\citenamefont {Urtizberea}\ \emph {et~al.}(2020)\citenamefont
  {Urtizberea}, \citenamefont {Natividad}, \citenamefont {Alonso},
  \citenamefont {P\'erez-Mart\'inez}, \citenamefont {Andr\'es}, \citenamefont
  {Gasc\'on}, \citenamefont {Gimeno}, \citenamefont {Luis},\ and\ \citenamefont
  {Roubeau}}]{Vanadyl3}%
  \BibitemOpen
  \bibfield  {author} {\bibinfo {author} {\bibfnamefont {A.}~\bibnamefont
  {Urtizberea}}, \bibinfo {author} {\bibfnamefont {E.}~\bibnamefont
  {Natividad}}, \bibinfo {author} {\bibfnamefont {P.~J.}\ \bibnamefont
  {Alonso}}, \bibinfo {author} {\bibfnamefont {L.}~\bibnamefont
  {P\'erez-Mart\'inez}}, \bibinfo {author} {\bibfnamefont {M.~A.}\ \bibnamefont
  {Andr\'es}}, \bibinfo {author} {\bibfnamefont {I.}~\bibnamefont {Gasc\'on}},
  \bibinfo {author} {\bibfnamefont {I.}~\bibnamefont {Gimeno}}, \bibinfo
  {author} {\bibfnamefont {F.}~\bibnamefont {Luis}},\ and\ \bibinfo {author}
  {\bibfnamefont {O.}~\bibnamefont {Roubeau}},\ }\href
  {https://doi.org/10.1039/C9MH01594A} {\bibfield  {journal} {\bibinfo
  {journal} {Mater. Horiz.}\ }\textbf {\bibinfo {volume} {7}},\ \bibinfo
  {pages} {885} (\bibinfo {year} {2020})}\BibitemShut {NoStop}%
\bibitem [{\citenamefont {Aguil\`a}\ \emph {et~al.}(2014)\citenamefont
  {Aguil\`a}, \citenamefont {Barrios}, \citenamefont {Velasco}, \citenamefont
  {Roubeau}, \citenamefont {Repoll\'es}, \citenamefont {Alonso}, \citenamefont
  {Ses\'e}, \citenamefont {Teat}, \citenamefont {Luis},\ and\ \citenamefont
  {Arom\'i}}]{Lanthanides1}%
  \BibitemOpen
  \bibfield  {author} {\bibinfo {author} {\bibfnamefont {D.}~\bibnamefont
  {Aguil\`a}}, \bibinfo {author} {\bibfnamefont {L.}~\bibnamefont {Barrios}},
  \bibinfo {author} {\bibfnamefont {V.}~\bibnamefont {Velasco}}, \bibinfo
  {author} {\bibfnamefont {O.}~\bibnamefont {Roubeau}}, \bibinfo {author}
  {\bibfnamefont {A.}~\bibnamefont {Repoll\'es}}, \bibinfo {author}
  {\bibfnamefont {P.~J.}\ \bibnamefont {Alonso}}, \bibinfo {author}
  {\bibfnamefont {J.}~\bibnamefont {Ses\'e}}, \bibinfo {author} {\bibfnamefont
  {S.~J.}\ \bibnamefont {Teat}}, \bibinfo {author} {\bibfnamefont
  {F.}~\bibnamefont {Luis}},\ and\ \bibinfo {author} {\bibfnamefont
  {G.}~\bibnamefont {Arom\'i}},\ }\bibfield  {booktitle} {\emph {\bibinfo
  {booktitle} {Journal of the American Chemical Society}},\ }\href
  {https://doi.org/10.1021/ja507809w} {\bibfield  {journal} {\bibinfo
  {journal} {Journal of the American Chemical Society}\ }\textbf {\bibinfo
  {volume} {136}},\ \bibinfo {pages} {14215} (\bibinfo {year}
  {2014})}\BibitemShut {NoStop}%
\bibitem [{\citenamefont {Aromí}\ and\ \citenamefont
  {Roubeau}(2019)}]{Lanthanides2}%
  \BibitemOpen
  \bibfield  {author} {\bibinfo {author} {\bibfnamefont {G.}~\bibnamefont
  {Aromí}}\ and\ \bibinfo {author} {\bibfnamefont {O.}~\bibnamefont
  {Roubeau}},\ }in\ \href
  {https://doi.org/https://doi.org/10.1016/bs.hpcre.2019.07.002} {\emph
  {\bibinfo {booktitle} {Including Actinides}}},\ \bibinfo {series} {Handbook
  on the Physics and Chemistry of Rare Earths}, Vol.~\bibinfo {volume} {56},\
  \bibinfo {editor} {edited by\ \bibinfo {editor} {\bibfnamefont {J.-C.~G.}\
  \bibnamefont {Bünzli}}\ and\ \bibinfo {editor} {\bibfnamefont {V.~K.}\
  \bibnamefont {Pecharsky}}}\ (\bibinfo  {publisher} {Elsevier},\ \bibinfo
  {year} {2019})\ pp.\ \bibinfo {pages} {1--54}\BibitemShut {NoStop}%
\bibitem [{\citenamefont {Carretta}\ \emph {et~al.}(2021)\citenamefont
  {Carretta}, \citenamefont {Zueco}, \citenamefont {Chiesa}, \citenamefont
  {G\'omez-Le\'on},\ and\ \citenamefont {Luis}}]{Perspective}%
  \BibitemOpen
  \bibfield  {author} {\bibinfo {author} {\bibfnamefont {S.}~\bibnamefont
  {Carretta}}, \bibinfo {author} {\bibfnamefont {D.}~\bibnamefont {Zueco}},
  \bibinfo {author} {\bibfnamefont {A.}~\bibnamefont {Chiesa}}, \bibinfo
  {author} {\bibfnamefont {A.}~\bibnamefont {G\'omez-Le\'on}},\ and\ \bibinfo
  {author} {\bibfnamefont {F.}~\bibnamefont {Luis}},\ }\href
  {https://doi.org/10.1063/5.0053378} {\bibfield  {journal} {\bibinfo
  {journal} {Applied Physics Letters}\ }\textbf {\bibinfo {volume} {118}},\
  \bibinfo {pages} {240501} (\bibinfo {year} {2021})},\ \Eprint
  {https://arxiv.org/abs/https://doi.org/10.1063/5.0053378}
  {https://doi.org/10.1063/5.0053378} \BibitemShut {NoStop}%
\bibitem [{\citenamefont {Kubo}\ \emph {et~al.}(2010)\citenamefont {Kubo},
  \citenamefont {Ong}, \citenamefont {Bertet}, \citenamefont {Vion},
  \citenamefont {Jacques}, \citenamefont {Zheng}, \citenamefont {Dr\'eau},
  \citenamefont {Roch}, \citenamefont {Auffeves}, \citenamefont {Jelezko},
  \citenamefont {Wrachtrup}, \citenamefont {Barthe}, \citenamefont {Bergonzo},\
  and\ \citenamefont {Esteve}}]{Strong-coupling0}%
  \BibitemOpen
  \bibfield  {author} {\bibinfo {author} {\bibfnamefont {Y.}~\bibnamefont
  {Kubo}}, \bibinfo {author} {\bibfnamefont {F.~R.}\ \bibnamefont {Ong}},
  \bibinfo {author} {\bibfnamefont {P.}~\bibnamefont {Bertet}}, \bibinfo
  {author} {\bibfnamefont {D.}~\bibnamefont {Vion}}, \bibinfo {author}
  {\bibfnamefont {V.}~\bibnamefont {Jacques}}, \bibinfo {author} {\bibfnamefont
  {D.}~\bibnamefont {Zheng}}, \bibinfo {author} {\bibfnamefont
  {A.}~\bibnamefont {Dr\'eau}}, \bibinfo {author} {\bibfnamefont {J.-F.}\
  \bibnamefont {Roch}}, \bibinfo {author} {\bibfnamefont {A.}~\bibnamefont
  {Auffeves}}, \bibinfo {author} {\bibfnamefont {F.}~\bibnamefont {Jelezko}},
  \bibinfo {author} {\bibfnamefont {J.}~\bibnamefont {Wrachtrup}}, \bibinfo
  {author} {\bibfnamefont {M.~F.}\ \bibnamefont {Barthe}}, \bibinfo {author}
  {\bibfnamefont {P.}~\bibnamefont {Bergonzo}},\ and\ \bibinfo {author}
  {\bibfnamefont {D.}~\bibnamefont {Esteve}},\ }\href
  {https://doi.org/10.1103/PhysRevLett.105.140502} {\bibfield  {journal}
  {\bibinfo  {journal} {Phys. Rev. Lett.}\ }\textbf {\bibinfo {volume} {105}},\
  \bibinfo {pages} {140502} (\bibinfo {year} {2010})}\BibitemShut {NoStop}%
\bibitem [{\citenamefont {Schuster}\ \emph {et~al.}(2010)\citenamefont
  {Schuster}, \citenamefont {Sears}, \citenamefont {Ginossar}, \citenamefont
  {DiCarlo}, \citenamefont {Frunzio}, \citenamefont {Morton}, \citenamefont
  {Wu}, \citenamefont {Briggs}, \citenamefont {Buckley}, \citenamefont
  {Awschalom},\ and\ \citenamefont {Schoelkopf}}]{Strong-coupling1}%
  \BibitemOpen
  \bibfield  {author} {\bibinfo {author} {\bibfnamefont {D.~I.}\ \bibnamefont
  {Schuster}}, \bibinfo {author} {\bibfnamefont {A.~P.}\ \bibnamefont {Sears}},
  \bibinfo {author} {\bibfnamefont {E.}~\bibnamefont {Ginossar}}, \bibinfo
  {author} {\bibfnamefont {L.}~\bibnamefont {DiCarlo}}, \bibinfo {author}
  {\bibfnamefont {L.}~\bibnamefont {Frunzio}}, \bibinfo {author} {\bibfnamefont
  {J.~J.~L.}\ \bibnamefont {Morton}}, \bibinfo {author} {\bibfnamefont
  {H.}~\bibnamefont {Wu}}, \bibinfo {author} {\bibfnamefont {G.~A.~D.}\
  \bibnamefont {Briggs}}, \bibinfo {author} {\bibfnamefont {B.~B.}\
  \bibnamefont {Buckley}}, \bibinfo {author} {\bibfnamefont {D.~D.}\
  \bibnamefont {Awschalom}},\ and\ \bibinfo {author} {\bibfnamefont {R.~J.}\
  \bibnamefont {Schoelkopf}},\ }\href
  {https://doi.org/10.1103/PhysRevLett.105.140501} {\bibfield  {journal}
  {\bibinfo  {journal} {Phys. Rev. Lett.}\ }\textbf {\bibinfo {volume} {105}},\
  \bibinfo {pages} {140501} (\bibinfo {year} {2010})}\BibitemShut {NoStop}%
\bibitem [{\citenamefont {Ams\"uss}\ \emph {et~al.}(2011)\citenamefont
  {Ams\"uss}, \citenamefont {Koller}, \citenamefont {N\"obauer}, \citenamefont
  {Putz}, \citenamefont {Rotter}, \citenamefont {Sandner}, \citenamefont
  {Schneider}, \citenamefont {Schramb\"ock}, \citenamefont {Steinhauser},
  \citenamefont {Ritsch}, \citenamefont {Schmiedmayer},\ and\ \citenamefont
  {Majer}}]{Strong-coupling2}%
  \BibitemOpen
  \bibfield  {author} {\bibinfo {author} {\bibfnamefont {R.}~\bibnamefont
  {Ams\"uss}}, \bibinfo {author} {\bibfnamefont {C.}~\bibnamefont {Koller}},
  \bibinfo {author} {\bibfnamefont {T.}~\bibnamefont {N\"obauer}}, \bibinfo
  {author} {\bibfnamefont {S.}~\bibnamefont {Putz}}, \bibinfo {author}
  {\bibfnamefont {S.}~\bibnamefont {Rotter}}, \bibinfo {author} {\bibfnamefont
  {K.}~\bibnamefont {Sandner}}, \bibinfo {author} {\bibfnamefont
  {S.}~\bibnamefont {Schneider}}, \bibinfo {author} {\bibfnamefont
  {M.}~\bibnamefont {Schramb\"ock}}, \bibinfo {author} {\bibfnamefont
  {G.}~\bibnamefont {Steinhauser}}, \bibinfo {author} {\bibfnamefont
  {H.}~\bibnamefont {Ritsch}}, \bibinfo {author} {\bibfnamefont
  {J.}~\bibnamefont {Schmiedmayer}},\ and\ \bibinfo {author} {\bibfnamefont
  {J.}~\bibnamefont {Majer}},\ }\href
  {https://doi.org/10.1103/PhysRevLett.107.060502} {\bibfield  {journal}
  {\bibinfo  {journal} {Phys. Rev. Lett.}\ }\textbf {\bibinfo {volume} {107}},\
  \bibinfo {pages} {060502} (\bibinfo {year} {2011})}\BibitemShut {NoStop}%
\bibitem [{\citenamefont {Bushev}\ \emph {et~al.}(2011)\citenamefont {Bushev},
  \citenamefont {Feofanov}, \citenamefont {Rotzinger}, \citenamefont
  {Protopopov}, \citenamefont {Cole}, \citenamefont {Wilson}, \citenamefont
  {Fischer}, \citenamefont {Lukashenko},\ and\ \citenamefont
  {Ustinov}}]{Strong-coupling-Spin0}%
  \BibitemOpen
  \bibfield  {author} {\bibinfo {author} {\bibfnamefont {P.}~\bibnamefont
  {Bushev}}, \bibinfo {author} {\bibfnamefont {A.~K.}\ \bibnamefont
  {Feofanov}}, \bibinfo {author} {\bibfnamefont {H.}~\bibnamefont {Rotzinger}},
  \bibinfo {author} {\bibfnamefont {I.}~\bibnamefont {Protopopov}}, \bibinfo
  {author} {\bibfnamefont {J.~H.}\ \bibnamefont {Cole}}, \bibinfo {author}
  {\bibfnamefont {C.~M.}\ \bibnamefont {Wilson}}, \bibinfo {author}
  {\bibfnamefont {G.}~\bibnamefont {Fischer}}, \bibinfo {author} {\bibfnamefont
  {A.}~\bibnamefont {Lukashenko}},\ and\ \bibinfo {author} {\bibfnamefont
  {A.~V.}\ \bibnamefont {Ustinov}},\ }\href
  {https://doi.org/10.1103/PhysRevB.84.060501} {\bibfield  {journal} {\bibinfo
  {journal} {Phys. Rev. B}\ }\textbf {\bibinfo {volume} {84}},\ \bibinfo
  {pages} {060501} (\bibinfo {year} {2011})}\BibitemShut {NoStop}%
\bibitem [{\citenamefont {Probst}\ \emph {et~al.}(2014)\citenamefont {Probst},
  \citenamefont {Tkal\ifmmode~\check{c}\else \v{c}\fi{}ec}, \citenamefont
  {Rotzinger}, \citenamefont {Rieger}, \citenamefont {Le~Floch}, \citenamefont
  {Goryachev}, \citenamefont {Tobar}, \citenamefont {Ustinov},\ and\
  \citenamefont {Bushev}}]{Strong-coupling-Spin1}%
  \BibitemOpen
  \bibfield  {author} {\bibinfo {author} {\bibfnamefont {S.}~\bibnamefont
  {Probst}}, \bibinfo {author} {\bibfnamefont {A.}~\bibnamefont
  {Tkal\ifmmode~\check{c}\else \v{c}\fi{}ec}}, \bibinfo {author} {\bibfnamefont
  {H.}~\bibnamefont {Rotzinger}}, \bibinfo {author} {\bibfnamefont
  {D.}~\bibnamefont {Rieger}}, \bibinfo {author} {\bibfnamefont {J.-M.}\
  \bibnamefont {Le~Floch}}, \bibinfo {author} {\bibfnamefont {M.}~\bibnamefont
  {Goryachev}}, \bibinfo {author} {\bibfnamefont {M.~E.}\ \bibnamefont
  {Tobar}}, \bibinfo {author} {\bibfnamefont {A.~V.}\ \bibnamefont {Ustinov}},\
  and\ \bibinfo {author} {\bibfnamefont {P.~A.}\ \bibnamefont {Bushev}},\
  }\href {https://doi.org/10.1103/PhysRevB.90.100404} {\bibfield  {journal}
  {\bibinfo  {journal} {Phys. Rev. B}\ }\textbf {\bibinfo {volume} {90}},\
  \bibinfo {pages} {100404} (\bibinfo {year} {2014})}\BibitemShut {NoStop}%
\bibitem [{\citenamefont {Bonizzoni}\ \emph {et~al.}(2017)\citenamefont
  {Bonizzoni}, \citenamefont {Ghirri}, \citenamefont {Atzori}, \citenamefont
  {Sorace}, \citenamefont {Sessoli},\ and\ \citenamefont
  {Affronte}}]{Strong-coupling-Spin2}%
  \BibitemOpen
  \bibfield  {author} {\bibinfo {author} {\bibfnamefont {C.}~\bibnamefont
  {Bonizzoni}}, \bibinfo {author} {\bibfnamefont {A.}~\bibnamefont {Ghirri}},
  \bibinfo {author} {\bibfnamefont {M.}~\bibnamefont {Atzori}}, \bibinfo
  {author} {\bibfnamefont {L.}~\bibnamefont {Sorace}}, \bibinfo {author}
  {\bibfnamefont {R.}~\bibnamefont {Sessoli}},\ and\ \bibinfo {author}
  {\bibfnamefont {M.}~\bibnamefont {Affronte}},\ }\href@noop {} {\bibfield
  {journal} {\bibinfo  {journal} {Scientific Reports}\ }\textbf {\bibinfo
  {volume} {7}},\ \bibinfo {pages} {13096} (\bibinfo {year}
  {2017})}\BibitemShut {NoStop}%
\bibitem [{\citenamefont {Mi}\ \emph {et~al.}(2018)\citenamefont {Mi},
  \citenamefont {Benito}, \citenamefont {Putz}, \citenamefont {Zajac},
  \citenamefont {Taylor}, \citenamefont {Burkard},\ and\ \citenamefont
  {Petta}}]{Spin-Photon0}%
  \BibitemOpen
  \bibfield  {author} {\bibinfo {author} {\bibfnamefont {X.}~\bibnamefont
  {Mi}}, \bibinfo {author} {\bibfnamefont {M.}~\bibnamefont {Benito}}, \bibinfo
  {author} {\bibfnamefont {S.}~\bibnamefont {Putz}}, \bibinfo {author}
  {\bibfnamefont {D.~M.}\ \bibnamefont {Zajac}}, \bibinfo {author}
  {\bibfnamefont {J.~M.}\ \bibnamefont {Taylor}}, \bibinfo {author}
  {\bibfnamefont {G.}~\bibnamefont {Burkard}},\ and\ \bibinfo {author}
  {\bibfnamefont {J.~R.}\ \bibnamefont {Petta}},\ }\href
  {https://doi.org/10.1038/nature25769} {\bibfield  {journal} {\bibinfo
  {journal} {Nature}\ }\textbf {\bibinfo {volume} {555}},\ \bibinfo {pages}
  {599} (\bibinfo {year} {2018})}\BibitemShut {NoStop}%
\bibitem [{\citenamefont {Samkharadze}\ \emph {et~al.}(2018)\citenamefont
  {Samkharadze}, \citenamefont {Zheng}, \citenamefont {Kalhor}, \citenamefont
  {Brousse}, \citenamefont {Sammak}, \citenamefont {Mendes}, \citenamefont
  {Blais}, \citenamefont {Scappucci},\ and\ \citenamefont
  {Vandersypen}}]{Vandersypen2018}%
  \BibitemOpen
  \bibfield  {author} {\bibinfo {author} {\bibfnamefont {N.}~\bibnamefont
  {Samkharadze}}, \bibinfo {author} {\bibfnamefont {G.}~\bibnamefont {Zheng}},
  \bibinfo {author} {\bibfnamefont {N.}~\bibnamefont {Kalhor}}, \bibinfo
  {author} {\bibfnamefont {D.}~\bibnamefont {Brousse}}, \bibinfo {author}
  {\bibfnamefont {A.}~\bibnamefont {Sammak}}, \bibinfo {author} {\bibfnamefont
  {U.~C.}\ \bibnamefont {Mendes}}, \bibinfo {author} {\bibfnamefont
  {A.}~\bibnamefont {Blais}}, \bibinfo {author} {\bibfnamefont
  {G.}~\bibnamefont {Scappucci}},\ and\ \bibinfo {author} {\bibfnamefont
  {L.~M.~K.}\ \bibnamefont {Vandersypen}},\ }\href
  {https://doi.org/10.1126/science.aar4054} {\bibfield  {journal} {\bibinfo
  {journal} {Science}\ }\textbf {\bibinfo {volume} {359}},\ \bibinfo {pages}
  {1123} (\bibinfo {year} {2018})}\BibitemShut {NoStop}%
\bibitem [{\citenamefont {Benito}\ \emph {et~al.}(2019)\citenamefont {Benito},
  \citenamefont {Petta},\ and\ \citenamefont {Burkard}}]{Benito2019}%
  \BibitemOpen
  \bibfield  {author} {\bibinfo {author} {\bibfnamefont {M.}~\bibnamefont
  {Benito}}, \bibinfo {author} {\bibfnamefont {J.~R.}\ \bibnamefont {Petta}},\
  and\ \bibinfo {author} {\bibfnamefont {G.}~\bibnamefont {Burkard}},\ }\href
  {https://doi.org/10.1103/PhysRevB.100.081412} {\bibfield  {journal} {\bibinfo
   {journal} {Phys. Rev. B}\ }\textbf {\bibinfo {volume} {100}},\ \bibinfo
  {pages} {081412} (\bibinfo {year} {2019})}\BibitemShut {NoStop}%
\bibitem [{\citenamefont {Gimeno}\ \emph {et~al.}(2020)\citenamefont {Gimeno},
  \citenamefont {Kersten}, \citenamefont {Pallarés}, \citenamefont
  {Hermosilla}, \citenamefont {Martínez-Pérez}, \citenamefont {Jenkins},
  \citenamefont {Angerer}, \citenamefont {Sánchez-Azqueta}, \citenamefont
  {Zueco}, \citenamefont {Majer}, \citenamefont {Lostao},\ and\ \citenamefont
  {Luis}}]{Constriction}%
  \BibitemOpen
  \bibfield  {author} {\bibinfo {author} {\bibfnamefont {I.}~\bibnamefont
  {Gimeno}}, \bibinfo {author} {\bibfnamefont {W.}~\bibnamefont {Kersten}},
  \bibinfo {author} {\bibfnamefont {M.~C.}\ \bibnamefont {Pallarés}}, \bibinfo
  {author} {\bibfnamefont {P.}~\bibnamefont {Hermosilla}}, \bibinfo {author}
  {\bibfnamefont {M.~J.}\ \bibnamefont {Martínez-Pérez}}, \bibinfo {author}
  {\bibfnamefont {M.~D.}\ \bibnamefont {Jenkins}}, \bibinfo {author}
  {\bibfnamefont {A.}~\bibnamefont {Angerer}}, \bibinfo {author} {\bibfnamefont
  {C.}~\bibnamefont {Sánchez-Azqueta}}, \bibinfo {author} {\bibfnamefont
  {D.}~\bibnamefont {Zueco}}, \bibinfo {author} {\bibfnamefont
  {J.}~\bibnamefont {Majer}}, \bibinfo {author} {\bibfnamefont
  {A.}~\bibnamefont {Lostao}},\ and\ \bibinfo {author} {\bibfnamefont
  {F.}~\bibnamefont {Luis}},\ }\href {https://doi.org/10.1021/acsnano.0c03167}
  {\bibfield  {journal} {\bibinfo  {journal} {ACS Nano}\ }\textbf {\bibinfo
  {volume} {14}},\ \bibinfo {pages} {8707} (\bibinfo {year} {2020})},\ \bibinfo
  {note} {pMID: 32441922},\ \Eprint
  {https://arxiv.org/abs/https://doi.org/10.1021/acsnano.0c03167}
  {https://doi.org/10.1021/acsnano.0c03167} \BibitemShut {NoStop}%
\bibitem [{\citenamefont {Borjans}\ \emph {et~al.}(2020)\citenamefont
  {Borjans}, \citenamefont {Croot}, \citenamefont {Mi}, \citenamefont
  {Gullans},\ and\ \citenamefont {Petta}}]{Borjans2020}%
  \BibitemOpen
  \bibfield  {author} {\bibinfo {author} {\bibfnamefont {F.}~\bibnamefont
  {Borjans}}, \bibinfo {author} {\bibfnamefont {X.~G.}\ \bibnamefont {Croot}},
  \bibinfo {author} {\bibfnamefont {X.}~\bibnamefont {Mi}}, \bibinfo {author}
  {\bibfnamefont {M.~J.}\ \bibnamefont {Gullans}},\ and\ \bibinfo {author}
  {\bibfnamefont {J.~R.}\ \bibnamefont {Petta}},\ }\href
  {https://doi.org/10.1038/s41586-019-1867-y} {\bibfield  {journal} {\bibinfo
  {journal} {Nature}\ }\textbf {\bibinfo {volume} {577}},\ \bibinfo {pages}
  {195} (\bibinfo {year} {2020})}\BibitemShut {NoStop}%
\bibitem [{\citenamefont {Harvey-Collard}\ \emph {et~al.}(2022)\citenamefont
  {Harvey-Collard}, \citenamefont {Dijkema}, \citenamefont {Zheng},
  \citenamefont {Sammak}, \citenamefont {Scappucci},\ and\ \citenamefont
  {Vandersypen}}]{Harvey-Collard2022}%
  \BibitemOpen
  \bibfield  {author} {\bibinfo {author} {\bibfnamefont {P.}~\bibnamefont
  {Harvey-Collard}}, \bibinfo {author} {\bibfnamefont {J.}~\bibnamefont
  {Dijkema}}, \bibinfo {author} {\bibfnamefont {G.}~\bibnamefont {Zheng}},
  \bibinfo {author} {\bibfnamefont {A.}~\bibnamefont {Sammak}}, \bibinfo
  {author} {\bibfnamefont {G.}~\bibnamefont {Scappucci}},\ and\ \bibinfo
  {author} {\bibfnamefont {L.~M.~K.}\ \bibnamefont {Vandersypen}},\ }\href
  {https://doi.org/10.1103/PhysRevX.12.021026} {\bibfield  {journal} {\bibinfo
  {journal} {Phys. Rev. X}\ }\textbf {\bibinfo {volume} {12}},\ \bibinfo
  {pages} {021026} (\bibinfo {year} {2022})}\BibitemShut {NoStop}%
\bibitem [{\citenamefont {Wang}\ \emph {et~al.}(2020)\citenamefont {Wang},
  \citenamefont {Hu}, \citenamefont {Sanders},\ and\ \citenamefont
  {Kais}}]{Qudits&QC}%
  \BibitemOpen
  \bibfield  {author} {\bibinfo {author} {\bibfnamefont {Y.}~\bibnamefont
  {Wang}}, \bibinfo {author} {\bibfnamefont {Z.}~\bibnamefont {Hu}}, \bibinfo
  {author} {\bibfnamefont {B.~C.}\ \bibnamefont {Sanders}},\ and\ \bibinfo
  {author} {\bibfnamefont {S.}~\bibnamefont {Kais}},\ }\href
  {https://doi.org/10.3389/fphy.2020.589504} {\bibfield  {journal} {\bibinfo
  {journal} {Frontiers in Physics}\ }\textbf {\bibinfo {volume} {8}},\ \bibinfo
  {pages} {479} (\bibinfo {year} {2020})}\BibitemShut {NoStop}%
\bibitem [{\citenamefont {Chiesa}\ \emph {et~al.}(2020)\citenamefont {Chiesa},
  \citenamefont {Macaluso}, \citenamefont {Petiziol}, \citenamefont
  {Wimberger}, \citenamefont {Santini},\ and\ \citenamefont
  {Carretta}}]{ErrorCorrection0}%
  \BibitemOpen
  \bibfield  {author} {\bibinfo {author} {\bibfnamefont {A.}~\bibnamefont
  {Chiesa}}, \bibinfo {author} {\bibfnamefont {E.}~\bibnamefont {Macaluso}},
  \bibinfo {author} {\bibfnamefont {F.}~\bibnamefont {Petiziol}}, \bibinfo
  {author} {\bibfnamefont {S.}~\bibnamefont {Wimberger}}, \bibinfo {author}
  {\bibfnamefont {P.}~\bibnamefont {Santini}},\ and\ \bibinfo {author}
  {\bibfnamefont {S.}~\bibnamefont {Carretta}},\ }\bibfield  {booktitle} {\emph
  {\bibinfo {booktitle} {The Journal of Physical Chemistry Letters}},\ }\href
  {https://doi.org/10.1021/acs.jpclett.0c02213} {\bibfield  {journal} {\bibinfo
   {journal} {The Journal of Physical Chemistry Letters}\ }\textbf {\bibinfo
  {volume} {11}},\ \bibinfo {pages} {8610} (\bibinfo {year}
  {2020})}\BibitemShut {NoStop}%
\bibitem [{\citenamefont {Macaluso}\ \emph {et~al.}(2020)\citenamefont
  {Macaluso}, \citenamefont {Rub\'in}, \citenamefont {Aguilà}, \citenamefont
  {Chiesa}, \citenamefont {Barrios}, \citenamefont {Mart\'inez}, \citenamefont
  {Alonso}, \citenamefont {Roubeau}, \citenamefont {Luis}, \citenamefont
  {Arom\'i},\ and\ \citenamefont {Carretta}}]{Error-Correction1}%
  \BibitemOpen
  \bibfield  {author} {\bibinfo {author} {\bibfnamefont {E.}~\bibnamefont
  {Macaluso}}, \bibinfo {author} {\bibfnamefont {M.}~\bibnamefont {Rub\'in}},
  \bibinfo {author} {\bibfnamefont {D.}~\bibnamefont {Aguilà}}, \bibinfo
  {author} {\bibfnamefont {A.}~\bibnamefont {Chiesa}}, \bibinfo {author}
  {\bibfnamefont {L.~A.}\ \bibnamefont {Barrios}}, \bibinfo {author}
  {\bibfnamefont {J.~I.}\ \bibnamefont {Mart\'inez}}, \bibinfo {author}
  {\bibfnamefont {P.~J.}\ \bibnamefont {Alonso}}, \bibinfo {author}
  {\bibfnamefont {O.}~\bibnamefont {Roubeau}}, \bibinfo {author} {\bibfnamefont
  {F.}~\bibnamefont {Luis}}, \bibinfo {author} {\bibfnamefont {G.}~\bibnamefont
  {Arom\'i}},\ and\ \bibinfo {author} {\bibfnamefont {S.}~\bibnamefont
  {Carretta}},\ }\href {https://doi.org/10.1039/D0SC03107K} {\bibfield
  {journal} {\bibinfo  {journal} {Chem. Sci.}\ }\textbf {\bibinfo {volume}
  {11}},\ \bibinfo {pages} {10337} (\bibinfo {year} {2020})}\BibitemShut
  {NoStop}%
\bibitem [{\citenamefont {Lockyer}\ \emph {et~al.}(2021)\citenamefont
  {Lockyer}, \citenamefont {Chiesa}, \citenamefont {Timco}, \citenamefont
  {McInnes}, \citenamefont {Bennett}, \citenamefont {Vitorica-Yrezebal},
  \citenamefont {Carretta},\ and\ \citenamefont
  {Winpenny}}]{Error-Correction2}%
  \BibitemOpen
  \bibfield  {author} {\bibinfo {author} {\bibfnamefont {S.~J.}\ \bibnamefont
  {Lockyer}}, \bibinfo {author} {\bibfnamefont {A.}~\bibnamefont {Chiesa}},
  \bibinfo {author} {\bibfnamefont {G.~A.}\ \bibnamefont {Timco}}, \bibinfo
  {author} {\bibfnamefont {E.~J.~L.}\ \bibnamefont {McInnes}}, \bibinfo
  {author} {\bibfnamefont {T.~S.}\ \bibnamefont {Bennett}}, \bibinfo {author}
  {\bibfnamefont {I.~J.}\ \bibnamefont {Vitorica-Yrezebal}}, \bibinfo {author}
  {\bibfnamefont {S.}~\bibnamefont {Carretta}},\ and\ \bibinfo {author}
  {\bibfnamefont {R.~E.~P.}\ \bibnamefont {Winpenny}},\ }\href
  {https://doi.org/10.1039/D1SC01506K} {\bibfield  {journal} {\bibinfo
  {journal} {Chem. Sci.}\ }\textbf {\bibinfo {volume} {12}},\ \bibinfo {pages}
  {9104} (\bibinfo {year} {2021})}\BibitemShut {NoStop}%
\bibitem [{\citenamefont {Petiziol}\ \emph {et~al.}(2021)\citenamefont
  {Petiziol}, \citenamefont {Chiesa}, \citenamefont {Wimberger}, \citenamefont
  {Santini},\ and\ \citenamefont {Carretta}}]{Error-Correction3}%
  \BibitemOpen
  \bibfield  {author} {\bibinfo {author} {\bibfnamefont {F.}~\bibnamefont
  {Petiziol}}, \bibinfo {author} {\bibfnamefont {A.}~\bibnamefont {Chiesa}},
  \bibinfo {author} {\bibfnamefont {S.}~\bibnamefont {Wimberger}}, \bibinfo
  {author} {\bibfnamefont {P.}~\bibnamefont {Santini}},\ and\ \bibinfo {author}
  {\bibfnamefont {S.}~\bibnamefont {Carretta}},\ }\href
  {https://doi.org/10.1038/s41534-021-00466-3} {\bibfield  {journal} {\bibinfo
  {journal} {npj Quantum Information}\ }\textbf {\bibinfo {volume} {7}},\
  \bibinfo {pages} {133} (\bibinfo {year} {2021})}\BibitemShut {NoStop}%
\bibitem [{\citenamefont {Lanyon}\ \emph {et~al.}(2009)\citenamefont {Lanyon},
  \citenamefont {Barbieri}, \citenamefont {Almeida}, \citenamefont {Jennewein},
  \citenamefont {Ralph}, \citenamefont {Resch}, \citenamefont {Pryde},
  \citenamefont {O'Brien}, \citenamefont {Gilchrist},\ and\ \citenamefont
  {White}}]{Qudits0}%
  \BibitemOpen
  \bibfield  {author} {\bibinfo {author} {\bibfnamefont {B.~P.}\ \bibnamefont
  {Lanyon}}, \bibinfo {author} {\bibfnamefont {M.}~\bibnamefont {Barbieri}},
  \bibinfo {author} {\bibfnamefont {M.~P.}\ \bibnamefont {Almeida}}, \bibinfo
  {author} {\bibfnamefont {T.}~\bibnamefont {Jennewein}}, \bibinfo {author}
  {\bibfnamefont {T.~C.}\ \bibnamefont {Ralph}}, \bibinfo {author}
  {\bibfnamefont {K.~J.}\ \bibnamefont {Resch}}, \bibinfo {author}
  {\bibfnamefont {G.~J.}\ \bibnamefont {Pryde}}, \bibinfo {author}
  {\bibfnamefont {J.~L.}\ \bibnamefont {O'Brien}}, \bibinfo {author}
  {\bibfnamefont {A.}~\bibnamefont {Gilchrist}},\ and\ \bibinfo {author}
  {\bibfnamefont {A.~G.}\ \bibnamefont {White}},\ }\href
  {https://doi.org/10.1038/nphys1150} {\bibfield  {journal} {\bibinfo
  {journal} {Nature Physics}\ }\textbf {\bibinfo {volume} {5}},\ \bibinfo
  {pages} {134} (\bibinfo {year} {2009})}\BibitemShut {NoStop}%
\bibitem [{\citenamefont {Campbell}(2014)}]{Qudits1}%
  \BibitemOpen
  \bibfield  {author} {\bibinfo {author} {\bibfnamefont {E.~T.}\ \bibnamefont
  {Campbell}},\ }\href {https://doi.org/10.1103/PhysRevLett.113.230501}
  {\bibfield  {journal} {\bibinfo  {journal} {Phys. Rev. Lett.}\ }\textbf
  {\bibinfo {volume} {113}},\ \bibinfo {pages} {230501} (\bibinfo {year}
  {2014})}\BibitemShut {NoStop}%
\bibitem [{\citenamefont {Tacchino}\ \emph {et~al.}(2021)\citenamefont
  {Tacchino}, \citenamefont {Chiesa}, \citenamefont {Sessoli}, \citenamefont
  {Tavernelli},\ and\ \citenamefont {Carretta}}]{Qudits-LightMatter}%
  \BibitemOpen
  \bibfield  {author} {\bibinfo {author} {\bibfnamefont {F.}~\bibnamefont
  {Tacchino}}, \bibinfo {author} {\bibfnamefont {A.}~\bibnamefont {Chiesa}},
  \bibinfo {author} {\bibfnamefont {R.}~\bibnamefont {Sessoli}}, \bibinfo
  {author} {\bibfnamefont {I.}~\bibnamefont {Tavernelli}},\ and\ \bibinfo
  {author} {\bibfnamefont {S.}~\bibnamefont {Carretta}},\ }\href
  {https://doi.org/10.1039/D1TC00851J} {\bibfield  {journal} {\bibinfo
  {journal} {J. Mater. Chem. C}\ }\textbf {\bibinfo {volume} {9}},\ \bibinfo
  {pages} {10266} (\bibinfo {year} {2021})}\BibitemShut {NoStop}%
\bibitem [{\citenamefont {Vargas-Calderón}\ \emph {et~al.}(2021)\citenamefont
  {Vargas-Calderón}, \citenamefont {Parra-A.}, \citenamefont {Vinck-Posada},\
  and\ \citenamefont {González}}]{SalesmanProblem}%
  \BibitemOpen
  \bibfield  {author} {\bibinfo {author} {\bibfnamefont {V.}~\bibnamefont
  {Vargas-Calderón}}, \bibinfo {author} {\bibfnamefont {N.}~\bibnamefont
  {Parra-A.}}, \bibinfo {author} {\bibfnamefont {H.}~\bibnamefont
  {Vinck-Posada}},\ and\ \bibinfo {author} {\bibfnamefont {F.~A.}\ \bibnamefont
  {González}},\ }\href {https://doi.org/10.7566/JPSJ.90.114002} {\bibfield
  {journal} {\bibinfo  {journal} {Journal of the Physical Society of Japan}\
  }\textbf {\bibinfo {volume} {90}},\ \bibinfo {pages} {114002} (\bibinfo
  {year} {2021})},\ \Eprint
  {https://arxiv.org/abs/https://doi.org/10.7566/JPSJ.90.114002}
  {https://doi.org/10.7566/JPSJ.90.114002} \BibitemShut {NoStop}%
\bibitem [{\citenamefont {Jenkins}\ \emph {et~al.}(2017)\citenamefont
  {Jenkins}, \citenamefont {Duan}, \citenamefont {Diosdado}, \citenamefont
  {Garc\'ia-Ripoll}, \citenamefont {Gaita-Ari\~no}, \citenamefont
  {Gim\'{e}nez-Saiz}, \citenamefont {Alonso}, \citenamefont {Coronado},\ and\
  \citenamefont {Luis}}]{GdW-Molecule}%
  \BibitemOpen
  \bibfield  {author} {\bibinfo {author} {\bibfnamefont {M.~D.}\ \bibnamefont
  {Jenkins}}, \bibinfo {author} {\bibfnamefont {Y.}~\bibnamefont {Duan}},
  \bibinfo {author} {\bibfnamefont {B.}~\bibnamefont {Diosdado}}, \bibinfo
  {author} {\bibfnamefont {J.~J.}\ \bibnamefont {Garc\'ia-Ripoll}}, \bibinfo
  {author} {\bibfnamefont {A.}~\bibnamefont {Gaita-Ari\~no}}, \bibinfo {author}
  {\bibfnamefont {C.}~\bibnamefont {Gim\'{e}nez-Saiz}}, \bibinfo {author}
  {\bibfnamefont {P.~J.}\ \bibnamefont {Alonso}}, \bibinfo {author}
  {\bibfnamefont {E.}~\bibnamefont {Coronado}},\ and\ \bibinfo {author}
  {\bibfnamefont {F.}~\bibnamefont {Luis}},\ }\href
  {https://doi.org/10.1103/PhysRevB.95.064423} {\bibfield  {journal} {\bibinfo
  {journal} {Physical Review B}\ }\textbf {\bibinfo {volume} {95}},\ \bibinfo
  {pages} {064423} (\bibinfo {year} {2017})}\BibitemShut {NoStop}%
\bibitem [{\citenamefont {Asenjo-Garcia}\ \emph {et~al.}(2019)\citenamefont
  {Asenjo-Garcia}, \citenamefont {Kimble},\ and\ \citenamefont
  {Chang}}]{Asenjo-Garcia2019}%
  \BibitemOpen
  \bibfield  {author} {\bibinfo {author} {\bibfnamefont {A.}~\bibnamefont
  {Asenjo-Garcia}}, \bibinfo {author} {\bibfnamefont {H.~J.}\ \bibnamefont
  {Kimble}},\ and\ \bibinfo {author} {\bibfnamefont {D.~E.}\ \bibnamefont
  {Chang}},\ }\href {https://doi.org/10.1073/pnas.1911467116} {\bibfield
  {journal} {\bibinfo  {journal} {Proceedings of the National Academy of
  Sciences}\ }\textbf {\bibinfo {volume} {116}},\ \bibinfo {pages} {25503}
  (\bibinfo {year} {2019})},\ \Eprint {https://arxiv.org/abs/1906.02204}
  {1906.02204} \BibitemShut {NoStop}%
\bibitem [{\citenamefont {Gatteschi}\ \emph {et~al.}(2006)\citenamefont
  {Gatteschi}, \citenamefont {Sessoli},\ and\ \citenamefont
  {Villain}}]{MolecularSpins1}%
  \BibitemOpen
  \bibfield  {author} {\bibinfo {author} {\bibfnamefont {D.}~\bibnamefont
  {Gatteschi}}, \bibinfo {author} {\bibfnamefont {R.}~\bibnamefont {Sessoli}},\
  and\ \bibinfo {author} {\bibfnamefont {J.}~\bibnamefont {Villain}},\
  }\href@noop {} {\emph {\bibinfo {title} {Molecular nanomagnets, Vol. 5}}}\
  (\bibinfo  {publisher} {Oxford University Press, New York},\ \bibinfo {year}
  {2006})\BibitemShut {NoStop}%
\bibitem [{\citenamefont {Bartolom\'e}\ \emph {et~al.}(2016)\citenamefont
  {Bartolom\'e}, \citenamefont {Luis}, ,\ and\ \citenamefont
  {Fern\'andez}}]{MolecularSpins2}%
  \BibitemOpen
  \bibfield  {author} {\bibinfo {author} {\bibfnamefont {J.}~\bibnamefont
  {Bartolom\'e}}, \bibinfo {author} {\bibfnamefont {F.}~\bibnamefont {Luis}},
  ,\ and\ \bibinfo {author} {\bibfnamefont {J.~F.}\ \bibnamefont
  {Fern\'andez}},\ }\href@noop {} {\emph {\bibinfo {title} {Molecular Magnets:
  Physics and Applications}}}\ (\bibinfo  {publisher} {Springer, Berlin,
  Heildelberg},\ \bibinfo {year} {2016})\BibitemShut {NoStop}%
\bibitem [{\citenamefont {Jenkins}\ \emph {et~al.}(2014)\citenamefont
  {Jenkins}, \citenamefont {Naether}, \citenamefont {Ciria}, \citenamefont
  {Sesé}, \citenamefont {Atkinson}, \citenamefont {Sánchez-Azqueta},
  \citenamefont {Barco}, \citenamefont {Majer}, \citenamefont {Zueco},\ and\
  \citenamefont {Luis}}]{Generlized-Dicke}%
  \BibitemOpen
  \bibfield  {author} {\bibinfo {author} {\bibfnamefont {M.~D.}\ \bibnamefont
  {Jenkins}}, \bibinfo {author} {\bibfnamefont {U.}~\bibnamefont {Naether}},
  \bibinfo {author} {\bibfnamefont {M.}~\bibnamefont {Ciria}}, \bibinfo
  {author} {\bibfnamefont {J.}~\bibnamefont {Sesé}}, \bibinfo {author}
  {\bibfnamefont {J.}~\bibnamefont {Atkinson}}, \bibinfo {author}
  {\bibfnamefont {C.}~\bibnamefont {Sánchez-Azqueta}}, \bibinfo {author}
  {\bibfnamefont {E.~d.}\ \bibnamefont {Barco}}, \bibinfo {author}
  {\bibfnamefont {J.}~\bibnamefont {Majer}}, \bibinfo {author} {\bibfnamefont
  {D.}~\bibnamefont {Zueco}},\ and\ \bibinfo {author} {\bibfnamefont
  {F.}~\bibnamefont {Luis}},\ }\href {https://doi.org/10.1063/1.4899141}
  {\bibfield  {journal} {\bibinfo  {journal} {Applied Physics Letters}\
  }\textbf {\bibinfo {volume} {105}},\ \bibinfo {pages} {162601} (\bibinfo
  {year} {2014})},\ \Eprint
  {https://arxiv.org/abs/https://doi.org/10.1063/1.4899141}
  {https://doi.org/10.1063/1.4899141} \BibitemShut {NoStop}%
\bibitem [{\citenamefont {G\'omez-Le\'on}\ \emph {et~al.}(2022)\citenamefont
  {G\'omez-Le\'on}, \citenamefont {Luis},\ and\ \citenamefont
  {Zueco}}]{AGL-Spectroscopy}%
  \BibitemOpen
  \bibfield  {author} {\bibinfo {author} {\bibfnamefont {A.}~\bibnamefont
  {G\'omez-Le\'on}}, \bibinfo {author} {\bibfnamefont {F.}~\bibnamefont
  {Luis}},\ and\ \bibinfo {author} {\bibfnamefont {D.}~\bibnamefont {Zueco}},\
  }\href {https://doi.org/10.1103/PhysRevApplied.17.064030} {\bibfield
  {journal} {\bibinfo  {journal} {Phys. Rev. Applied}\ }\textbf {\bibinfo
  {volume} {17}},\ \bibinfo {pages} {064030} (\bibinfo {year}
  {2022})}\BibitemShut {NoStop}%
\bibitem [{\citenamefont {Schrieffer}\ and\ \citenamefont
  {Wolff}(1966)}]{Wolff1966}%
  \BibitemOpen
  \bibfield  {author} {\bibinfo {author} {\bibfnamefont {J.~R.}\ \bibnamefont
  {Schrieffer}}\ and\ \bibinfo {author} {\bibfnamefont {P.~A.}\ \bibnamefont
  {Wolff}},\ }\href {https://doi.org/10.1103/PhysRev.149.491} {\bibfield
  {journal} {\bibinfo  {journal} {Phys. Rev.}\ }\textbf {\bibinfo {volume}
  {149}},\ \bibinfo {pages} {491} (\bibinfo {year} {1966})}\BibitemShut
  {NoStop}%
\bibitem [{\citenamefont {Janowicz}(2003)}]{Multiple-scales-optics}%
  \BibitemOpen
  \bibfield  {author} {\bibinfo {author} {\bibfnamefont {M.}~\bibnamefont
  {Janowicz}},\ }\href
  {https://doi.org/https://doi.org/10.1016/S0370-1573(02)00551-3} {\bibfield
  {journal} {\bibinfo  {journal} {Physics Reports}\ }\textbf {\bibinfo {volume}
  {375}},\ \bibinfo {pages} {327} (\bibinfo {year} {2003})}\BibitemShut
  {NoStop}%
\bibitem [{\citenamefont {G\'omez-Le\'on}\ and\ \citenamefont
  {Platero}(2020)}]{PRR-bichromatic}%
  \BibitemOpen
  \bibfield  {author} {\bibinfo {author} {\bibfnamefont {A.}~\bibnamefont
  {G\'omez-Le\'on}}\ and\ \bibinfo {author} {\bibfnamefont {G.}~\bibnamefont
  {Platero}},\ }\href {https://doi.org/10.1103/PhysRevResearch.2.033412}
  {\bibfield  {journal} {\bibinfo  {journal} {Phys. Rev. Research}\ }\textbf
  {\bibinfo {volume} {2}},\ \bibinfo {pages} {033412} (\bibinfo {year}
  {2020})}\BibitemShut {NoStop}%
\bibitem [{\citenamefont {G\'omez-Le\'on}(2019)}]{Multiple-scales-SpinBath}%
  \BibitemOpen
  \bibfield  {author} {\bibinfo {author} {\bibfnamefont {A.}~\bibnamefont
  {G\'omez-Le\'on}},\ }\href {https://doi.org/10.1103/PhysRevB.100.094308}
  {\bibfield  {journal} {\bibinfo  {journal} {Phys. Rev. B}\ }\textbf {\bibinfo
  {volume} {100}},\ \bibinfo {pages} {094308} (\bibinfo {year}
  {2019})}\BibitemShut {NoStop}%
\bibitem [{Note1()}]{Note1}%
  \BibitemOpen
  \bibinfo {note} {Notice that the physical small parameter when $\epsilon \to
  1$ will be $\protect \tilde {J}_{i,j}^{\protect \vec {\alpha },\protect \vec
  {\beta }}$}\BibitemShut {NoStop}%
\bibitem [{\citenamefont {Doherty}\ \emph {et~al.}(2012)\citenamefont
  {Doherty}, \citenamefont {Dolde}, \citenamefont {Fedder}, \citenamefont
  {Jelezko}, \citenamefont {Wrachtrup}, \citenamefont {Manson},\ and\
  \citenamefont {Hollenberg}}]{NV-center}%
  \BibitemOpen
  \bibfield  {author} {\bibinfo {author} {\bibfnamefont {M.~W.}\ \bibnamefont
  {Doherty}}, \bibinfo {author} {\bibfnamefont {F.}~\bibnamefont {Dolde}},
  \bibinfo {author} {\bibfnamefont {H.}~\bibnamefont {Fedder}}, \bibinfo
  {author} {\bibfnamefont {F.}~\bibnamefont {Jelezko}}, \bibinfo {author}
  {\bibfnamefont {J.}~\bibnamefont {Wrachtrup}}, \bibinfo {author}
  {\bibfnamefont {N.~B.}\ \bibnamefont {Manson}},\ and\ \bibinfo {author}
  {\bibfnamefont {L.~C.~L.}\ \bibnamefont {Hollenberg}},\ }\href
  {https://doi.org/10.1103/PhysRevB.85.205203} {\bibfield  {journal} {\bibinfo
  {journal} {Phys. Rev. B}\ }\textbf {\bibinfo {volume} {85}},\ \bibinfo
  {pages} {205203} (\bibinfo {year} {2012})}\BibitemShut {NoStop}%
\bibitem [{\citenamefont {Rollano}\ \emph {et~al.}(2022)\citenamefont
  {Rollano}, \citenamefont {de~Ory}, \citenamefont {Buch}, \citenamefont
  {Rubín-Osanz}, \citenamefont {Zueco}, \citenamefont {Sánchez-Azqueta},
  \citenamefont {Chiesa}, \citenamefont {Granados}, \citenamefont {Carretta},
  \citenamefont {Gomez}, \citenamefont {Piligkos},\ and\ \citenamefont
  {Luis}}]{HighCooperativity}%
  \BibitemOpen
  \bibfield  {author} {\bibinfo {author} {\bibfnamefont {V.}~\bibnamefont
  {Rollano}}, \bibinfo {author} {\bibfnamefont {M.~C.}\ \bibnamefont {de~Ory}},
  \bibinfo {author} {\bibfnamefont {C.~D.}\ \bibnamefont {Buch}}, \bibinfo
  {author} {\bibfnamefont {M.}~\bibnamefont {Rubín-Osanz}}, \bibinfo {author}
  {\bibfnamefont {D.}~\bibnamefont {Zueco}}, \bibinfo {author} {\bibfnamefont
  {C.}~\bibnamefont {Sánchez-Azqueta}}, \bibinfo {author} {\bibfnamefont
  {A.}~\bibnamefont {Chiesa}}, \bibinfo {author} {\bibfnamefont
  {D.}~\bibnamefont {Granados}}, \bibinfo {author} {\bibfnamefont
  {S.}~\bibnamefont {Carretta}}, \bibinfo {author} {\bibfnamefont
  {A.}~\bibnamefont {Gomez}}, \bibinfo {author} {\bibfnamefont
  {S.}~\bibnamefont {Piligkos}},\ and\ \bibinfo {author} {\bibfnamefont
  {F.}~\bibnamefont {Luis}},\ }\href
  {https://doi.org/10.48550/ARXIV.2203.00965} {\bibinfo {title} {High
  cooperativity coupling to nuclear spins on a circuit qed architecture}}
  (\bibinfo {year} {2022})\BibitemShut {NoStop}%
\end{thebibliography}%
\end{document}